\documentclass[twocolumn]{aastex631}

\usepackage{natbib} 
\usepackage{color}
\usepackage{xspace}
\usepackage{amsmath}

\newcommand{\trat}{$T_{\rm p}/T_{\rm e}$\xspace}
\newcommand{\qratio}{$Q^+_{\rm p}/Q^+_{\rm e}$\xspace}

\newcommand\rhigh{R_{\rm high}}
\newcommand\rlow{R_{\rm low}}

\shorttitle{}
\shortauthors{}
\graphicspath{{./}{./}}
\begin{document}

\title{Revision of two-temperature magnetically arrested flows onto a black hole}

\correspondingauthor{M. Mo{\'s}cibrodzka}
\email{m.moscibrodzka@astro.ru.nl}
\author[0000-0002-4661-6332]{M. Mo{\'s}cibrodzka}
\affiliation{Department of Astrophysics/IMAPP, Radboud University,P.O. Box
  9010, 6500 GL Nijmegen, The Netherlands}

\begin{abstract}
We revisit the radiative properties of 3D general relativistic magnetohydrodynamics (GRMHD) two-temperature magnetically
arrested disk (MAD) models in which electrons are heated by a magnetic turbulent
cascade. We focus on studying the model emission, whose characteristics
include variability in both total intensity and linear/circular polarizations
as well as rotation measures at energies around the synchrotron emission peak in millimeter waves. We
find that the radiative properties of MAD models with turbulent electron heating are well converged with respect to the numerical grid resolution, which has not been demonstrated before. We compare radiation from two-temperature simulations with turbulent heating to
single-temperature models with electron temperatures calculated based on the 
commonly used $R~(\beta)$ prescription. We find that the
self-consisitent two-temperature models with
turbulent heating do not significantly
outperform the $R~(\beta)$ models and, in practice, may be indistinguishable
from the $R~(\beta)$ models. Accounting for physical effects such as radiative
cooling and the nonthermal electron distribution function makes a weak impact on properties of millimeter emission. Models are scaled to Sgr~A*, an accreting black hole in the center of our galaxy, and compared to the most complete observational datasets. We point out the consistencies and inconsistencies between the MAD models and observations of this source and discuss future prospects for GRMHD simulations. 
\end{abstract}

%% https://astrothesaurus.org
\keywords{Supermassive black holes (1663), Magnetohydrodynamical simulations (1966), 
Low-luminosity active galactic nuclei
(2033), Plasma physics (2089)} 

\section{Introduction}

Magnetically arrested disks (hereafter MADs) are a class of black hole accretion solutions
initially developed for
radiatively inefficient accretion flows (RIAFs; \citealt{Igumenshchev:2003,Narayan:2003,narayan:2022,proga:2003,Tchekhovskoy:2011,mckinney:2012}). In
MADs realized in general relativistic magnetohydrodynamics (GRMHD) simulations,
the magnetic field flux accumulated near the black hole event horizon is extreme, which has
two important consequences. First, the power of jets produced in MAD
simulations is significant, which makes the jet opening angle rather large ($\sim 55^\circ$; \citealt{chael:2019}). This is
consistent with what is observed in sub-Eddington accreting black hole systems
such as M87 \citep{kim:2018,walker:2018}. MADs near the
black hole horizon are further supported in M87 by the first
polarimetric images from the Event Horizon Telescope (EHT;
\citealt{EHTC:2021VIII}). Second, the jet magnetosphere in MADs is often
reconnecting at the equatorial plane of the simulation causing magnetic field
flux eruptions into the disk body. Such flux eruptions have been proposed to explain the flaring behavior of another sub-Eddington accreting
supermassive black hole, Sgr~A*
\citep{dexter:2020a,porth:2021,wielgus:2022b}. It has also been recently found
that horizon-scale polarimetric images of Sgr~A* are more consistent with MADs
\citep{EHTC:2024VIII}.
Although less magnetized standard and normal evolution accretion disks (SANEs), or wind-fed accretion or tilted-disk models (e.g., \citealt{ressler:2019,chatterejee:2020}), are not
completely ruled out by observations of any of the sources, the MAD RIAF solution
became a leading candidate to explain the observations of the two best-resolved
aforementioned accreting supermassive black holes.

One of the most uncertain parts of MADs (but also SANEs) is
how they generate emission. In the aforementioned class of systems, where the accreting plasma is
collisionless, electron and ion distribution functions remain unknown. Typically, to mimic the collisionless effects in models, we assume that the
plasma has two-temperature structure, and we ``paint'' electron temperatures on
the top of the GRMHD models according to some parameterized law.
An example of such a parameterized law for the thermal distribution of
electrons is the $R~(\beta$) prescription of \citealt{moscibrodzka:2016} used in
\citealt{EHTC:2022V}, where the model parameter $R~(\beta)\equiv T_{\rm i}/T_{\rm e}(\beta)$ is a ratio of
the nonemitting ion to the emitting electron temperatures that depends on the local
plasma $\beta(\equiv P_{\rm mag}/P_{\rm gas}$) parameter.
Other than the $R~(\beta)$ models has been proposed by \citet{moscibrodzka:2013,moscibrodzka:2014,chan:2015,gold:2017,anantua:2020}.
Alternatively, one can assume a nonthermal electron distribution function, but those are
typically also parametric (see, e.g., \citealt{davelaar:2018,cruzosorio:2022,fromm:2022,scepi:2022,zhao:2023} for recent work and \citealt{chael:2017,petersen:2020} for examples of models with evolution of nonthermal electrons). 
Finally, one can model two-temperature GRMHD flows where electron temperatures are followed by a
separate equation that includes nonadiabatic (viscous), subgrid heating terms based on plasma or
particle-in-cell considerations (\citealt{howes:2010,rowan:2017,kawazura:2019}) and
radiative cooling terms (e.g., \citealt{ryan:2018}). 

The method of tracking electron temperatures in
GRMHD simulations assuming a subgrid model for ion and electron heating has
been introduced by \citet{ressler:2015}. \citet{sadowski:2017} further
developed the idea to add the ion/electron pressure as well as a self-consistent variable adiabatic index to the GRMHD evolution. Two-temperature 2D and 3D GRMHD models have been carried out in the past, specifically for Sgr~A* (\citealt{ressler:2017,sadowski:2017,chael:2018,dexter:2020b,Jiang:2023}) and M87 (\citealt{ryan:2018,Mizuno:2021,Dihingia:2023}).

In this paper, we revisit the two-temperature GRMHD MAD simulations. The mentioned previous studies used several numerical codes assuming various torus sizes, magnetic field topologies, adiabatic indices, and grid resolutions to investigate the problem. It was not
clear to us that the radiative output of these models is converged for MAD models with setups comparable to those in the EHT simulation library \citep{EHTC:2022V}. The second motivation for the revision of two-temperature simulations is the fact that MAD models with $R~(\beta)$ prescription typically have too-variable total intensity compared to the observations of Sgr~A* \citep{EHTC:2022V,wielgus:2022a}. The origin of the discrepancy between the models and the observations is unknown. One possibility is that the $R~(\beta)$ model, which was initially developed for SANE simulations, is not suitable for MADs, and a more sophisticated electron model should be adopted. One could also ask how physically motivated the $R~(\beta)$ model is for electron temperatures when considering MADs. The latter is relevant for all sub-Eddington-accreting black hole systems. Finally, the measurement of black hole spin in EHT and other sources is typically accretion-model-dependent. A robust understanding of the dissipation processes in accreting plasma near the event horizon is critical for reliable spin estimates. 

In this work, we investigate how the numerical model parameters
impact the radiative (total intensity and polarimetric) characteristics of the two-temperature MAD simulations. 
We first demonstrate that they are
independent of the grid resolution used in GRMHD simulations.
Next, we evolve models for longer times and compare the two-temperature models with $R~(\beta)$ models. 
Finally, having checked that the results are weakly dependent on the exact
shape of the electron distribution function and that they remain unchanged
even when radiative cooling is introduced (scaling models to the Sgr~A* system),
we compare the prograde and retrograde two-temperature MAD models to the
newest multifrequency polarimetric millimeter observational data of Sgr~A*.
In contrast to most of the previous two-temperature simulations,
in our comparisons, we focus on variability and polarization,
which together are significantly more informative than the time-averaged or
total intensity comparisons. We discuss the future prospects for investigating the thermodynamics of the GRMHD simulations. 

The paper is structured as follows. In Section~\ref{sec:models} we describe
our 3D two-temperature GRMHD simulations setups and outline the details of radiative transfer modeling. In
Section~\ref{sec:results}, we report the results and compare the models with selected observations of Sgr~A*. We discuss the results and conclude in Section~\ref{sec:discussion}.

\section{Models}\label{sec:models}

\subsection{GRMHD simulations}

MAD simulations are carried by means of \texttt{ebhlight}, a radiative GRMHD code developed and made public by
\citet{ryan:2015}. \texttt{ebhlight} is a relativistic, second-order, conservative, constrained-transport code for stationary spacetimes. While the fluid part of the code is based on the \texttt{harm} routines \citep{gammie:2003}, the radiation part is based on \texttt{grmonty} scheme  \citep{dolence:2009}. 

All simulations are set in geometrized units where the
length scale and time scale are set by the black hole mass: ${\mathcal L} \equiv
GM/c^2$ and ${\mathcal T}\equiv GM/c^3$ (assuming $G=c=1$ both time and
distance are measured in units of mass, M).  The simulations start from the
Fishbone-Moncrief torus \citep{fishbone:1976} at the equatorial plane of a
Schwarzschild or Kerr black hole. The torus is described by two parameters:
inner radius $r_{\rm in}=20$M and radius of the pressure maximum
$r_{\rm max}=41$M. The torus is seeded with weak poloidal magnetic fields described by
vector potential,
\begin{equation}
(A_r,A_\theta,A_\phi)=(0,0, \frac{\rho}{\rho_{max}}  \left(\frac{r}{r_{in}}\right)^3\exp\left(\frac{r}{r_0}\right)\sin^3\theta -0.2)
\end{equation}  
where $r,\theta$ are the radius and polar angle in Kerr-Schild coordinates, parameter $r_0=400$M, $\rho$ is the plasma density. The initial magnetic field is
renormalized so that plasma $\beta_{\rm max}=100$.

The simulations use ideal equation of state with constant adiabatic index of $\gamma_{ad}=13/9$; i.e., we assume that plasma is pure hydrogen with nonrelativistic protons and relativistic electrons for which adiabatic indices are $\gamma_{\rm p}=5/3$ and $\gamma_{\rm e}=4/3$, respectively.

All simulations are carried out in mixed modified Kerr-Schield logarithmic
coordinates, where the resolution is focused on the equatorial plane and toward the
central region close to the black hole horizon. The grid stretches from within the black hole event horizon until $r_{\rm out}=1000$M. 

The grid refinement convergence test runs are integrated for $10,000$M
assuming four grid resolutions listed in Table~\ref{tab:all}.
Convergence runs are performed for black hole spin $a_*=0$. Later
in the paper, we consider both prograde and retrograde fiducial models (also
listed in Table~\ref{tab:all}) with the
following spin values: $a_*$=-0.9375, -0.5, 0, 0.5, 0.9375. The prograde and zero-spin
models are evolved until $30,000$M, and the retrograde models are
evolved for a shorter time, only until $14,000$M.
In all our simulations, the accretion flows reach steady state within
$r=20$M. Fiducial models have grid resolutions ($N_r,N_\theta,N_\phi$) = (240,
120, 128) (for
$a_*$ = -0.5, 0, 0.5) and ($N_r,N_\theta,N_\phi$) = (266, 120, 128) (for
$a_*$ = -0.9375, 0.9375). All models above are run with radiative modules of
\texttt{ebhlight} turned off. We carry out one test run with radiative effects turned on. The
parameters of the exploratory radiative GRMHD (GRRMHD) model are shown in Table~\ref{tab:all}.

\subsection{Radiative transfer}

We measure the convergence of the models by studying their radiative properties.
To predict synchrotron emission from the simulations, we postprocessed the
GRMHD snapshots using the ray-tracing
relativistic polarized radiative transfer code \texttt{ipole}
\citep{moscibrodzka:2018}. The radiative transfer calculations are typically carried out starting at later times of simulations when the accretion flow is relaxed from the initial conditions (starting times, $t_{\rm s}$, and final times, $t_{\rm f}$, of all radiative transfer postprocessing simulations are listed in Table~\ref{tab:all}).
We scale all GRMHD simulations using the mass and distance of the Sgr~A* black
hole. The models density scale is set by unit ${\mathcal M}$, a standard mass or
accretion rate $\dot{M}$ scaling factor (e.g., 
$\rho_{Sgr~A*}={\mathcal M} / {\mathcal L}^3 \rho_{code}$). 
In all models $\mathcal M$ is set to produce an average Sgr~A* total flux of 2-3
Jy observed at a frequency of 229 GHz \citep{wielgus:2022a}. 
Table~\ref{tab:all} lists each model ${\mathcal M}$ and corresponding $\dot{M}$. Given the electron
temperature model (see next subsection), the imaging code produces synchrotron emission
maps at a desired frequency $\nu$ (here 86--229 GHz) and a viewing angle $i$.
Initially, all maps are computed for a default viewing angle $i=160^\circ$
selected based on the recent rediscovery of a transient hot spot orbiting around
Sgr~A* (\citealt{wielgus:2022b}, see also viewing angle estimates by
\citealt{yfantis:2024})
and then later for additional viewing angles $i=150^\circ, 130^\circ,
110^\circ$. Note that for the modeled source, lower viewing angles are favored
by EHT observations that revealed a symmetric ring \citep{EHTC:2022V}; hence we do not model emission for high viewing angles, $i\sim 90^\circ$. Assuming $i>90^\circ$ also guarantees the correct sign of the circular polarization for the assumed polarity of magnetic fields in the GRMHD models. 
Our model map resolution is $256 \times 256$ pixels and the field of view =
$120$M $\approx 600$ $\mu as$. A time series of maps/images with a cadence of $\Delta
t=10$M is synthesized into light curves in Stokes ${\mathcal
  I},{\mathcal Q},{\mathcal U},{\mathcal V}$. 
Following \citealt{wielgus:2022a} the modulation index
$M_3{}\equiv\sigma_{\Delta T}/\mu_{\Delta T}$ (ratio of standard deviation to
mean flux calculated on time intervals $\Delta T$ = 3 hr) is used to
characterize light-curve variability in Stokes ${\mathcal I}$ at two
frequencies, 86 and 229 GHz. 
We also calculate the spectral index in total intensity ($\alpha_I$ between
213 and 229 GHz), (Faraday) rotation
measure (RM; between 213 and 229 GHz), and linear (LP) and circular (CP) fractional polarizations at 86 and 229~GHz. These particular quantities are studied because they are later compared directly to observations. 
In our models Faraday rotation is caused by relativistic and subrelativistic
electrons within 100M from the event horizon; these electrons constitute the so-called internal Faraday screen. 

\subsection{Electron distribution functions}\label{sec:electrons}

Following the scheme of \citet{ressler:2015}, the \texttt{ebhlight} code tracks total and electron entropies that are used to calculate the nonadiabatic (viscous) gas heating rates and evolve electron temperatures. The scheme requires the provision of a subgrid model for the proton-to-electron heating ratio \qratio. In this work, we adopt
the prescription for \qratio developed by \citet[][henceforth model K]{kawazura:2019}, which approximates dissipation in a turbulent cascade with a functional form
\begin{equation}
    \frac{Q^+_{\rm{p}}}{Q^+_{\rm{e}}} = \frac{35}{1+(\beta/15)^{-1.4}\exp^{-0.1T_e/T_p}}.
\end{equation}
In this model, most of the dissipation goes to protons in high $\beta$ plasma
regions, while electrons receive most of the heating in low $\beta$
regions. Qualitatively and quantitatively, the K model is very similar to the turbulent heating model of \citet{howes:2010}. Studying significantly different options for \qratio, such as, for example, electron heating by reconnection \citep{rowan:2017}, is beyond the scope of the current paper but is discussed in Section~\ref{sec:discussion}.

In ideal GRMHD models studied here using a conservative code, the total viscous heating is produced
by truncation errors at the numerical grid level. One may conclude that no
matter what fraction of this heating goes to the electrons, this heating will be
completely artificial. However, for the turbulent torus problem, the
grid-scale dissipation is set by the large-scale turbulence in
the problem (see Section 3.1 in \citealt{ressler:2015} for a more detailed discussion). The numerical scheme for electron heating implemented in \texttt{ebhlight} has been carefully tested against some analytic problems by \citet{ressler:2015} and \citet{ryan:2017} as well as \citet{sadowski:2017}, who, except for the mentioned modifications, follow the same idea when evolving electron temperatures alone. We are not going to repeat these tests here.  
In \texttt{ebhlight} the electron evolution also does not include Coulomb couplings, unless radiative transfer is activated (see Section~\ref{section:cool}). However, those are not important in the low-density gas considered here.
In \texttt{ebhlight} the electron temperature tracking is passive; i.e. the
electron pressure is not accounted for in the GRMHD equations. More
importantly, the heating of the proton (nonadiabatic/viscous) is not taken into account in the GRMHD equations as well; instead, the internal energy of the gas is evaluated from the total energy using an inversion scheme implemented in the code assuming $\gamma_{ad}$. The proton temperatures are computed using the total internal energy $u$, the plasma density $\rho$, and $\gamma_{\rm p}=5/3$ ($\Theta_{\rm p}\equiv (\gamma_{\rm p}-1)u/\rho$).

The simulations are two-temperature, but to make a connection with the previous
studies, we use the same GRMHD runs and compute images/light curves using a parametric electron temperature
model $R~(\beta)$. In this model, proton
temperatures are calculated from GRMHD quantities, and the electron temperature is
found using the formula below:
\begin{equation}
R(\beta) \equiv \frac{T_{\rm{p}}}{T_{\rm{e}}} = R_{\rm{high}} \frac{\beta^2}{1+\beta^2}+R_{\rm{low}} \frac{1}{1+\beta^2}, 
    \label{eq:rhigh}
\end{equation}
where parameters $\rlow$ and $\rhigh$ are temperature ratios that describe the
proton-to-electron temperature ratio in strongly (low plasma $\beta$) and weakly (high
plasma $\beta$) magnetized regions, respectively. To be consistent with
\citet{EHTC:2022V,EHTC:2024VIII}, we calculate the light curves for $\rhigh =
1, 10, 40, 160$ and
$\rlow = 1$. The dimensionless electron temperature that is passed to the synchrotron
transfer coefficients in \texttt{ipole} is calculated from
\begin{equation}\label{eq:thetae}
\Theta_e = \frac{u}{\rho} \frac{m_{\rm p}}{m_{\rm e}} \frac{(\gamma_{\rm
    p}-1)(\gamma_{\rm e}-1)}{(\gamma_{\rm e}-1)R + (\gamma_{\rm p}-1)}
\end{equation}  
with internal energy $u$ and rest-mass density $\rho$ provided by GRMHD model. The electron model
assumes $\gamma_{\rm e} = 4/3$ and $\gamma_{\rm p} = 5/3$, self-consistent with
GRMHD simulations in which electrons are relativistic and protons are nonrelativistic.  

Models with the \citet{kawazura:2019}
prescription for \qratio are tagged K, and models with
parametric electron temperatures are labeled R1-R160 (e.g., model with $\rlow=1$
and $\rhigh=160$ is denoted R160). The latter models are referred to as $R~(\beta)$ models.

It is reasonable to assume that the energy from the turbulent heating is used to form a nonthermal rather than purely thermal electron distribution function. We consider an exploratory 
nonthermal model using the $f_\kappa(\gamma,\kappa,w)$ distribution function
\citep{xiao:2006}:
\begin{equation}
f_\kappa(\gamma,\kappa,w)=N \gamma \sqrt{\gamma^2-1} \left(1+\frac{\gamma-1}{\kappa w}\right)^{-(\kappa+1)}
\end{equation}
where $\kappa$ and $w$ are parameters and $N$ is the normalization constant. 
The distribution parameter
\begin{equation}\label{eq:kappa}
w=\frac{(\kappa-3)}{\kappa} \Theta_e,
\end{equation}
where $\Theta_e$ is calculated form the electron entropy in exactly the same
manner as in purely thermal K models (see \citealt{ressler:2015}).
Equation~\ref{eq:kappa} states that the entire turbulent energy is distributed
into the $\kappa$ function. 
The parameter $\kappa$ is globally constant. Inspired by the solar wind
studies, we assume $\kappa = 4.25$ \citep[e.g.,][]{liv:2018}. In ray-tracing, the radiative transfer coefficients for the
$\kappa$ distribution function are adopted from \citet{moscibrodzka:2024}. No
cutoff is applied to the $\kappa$ distribution function at high energies because it makes a negligible difference when modeling (sub)millimeter emission.

All radiative transfer simulations are carried out only in regions where
$\sigma < \sigma_{\rm cut} = 1$ (where magnetization parameter $\sigma \equiv 2P_{\rm mag}/\rho c^2$). This is a standard procedure to avoid modeling
emission from near-vacuum jet regions in the GRMHD model where the proton and electron
temperatures are inaccurate \citep[e.g.,][]{EHTC:2022V}.
However, the millimeter emission from regions with $\sigma > 1$ is small, as
already shown by \citet[][see their Appendix D]{dexter:2020b}.

\subsection{Effects of grid resolution}\label{sec:convergence}

\begin{figure*}
  \centering
   \includegraphics[width=1.0\linewidth]{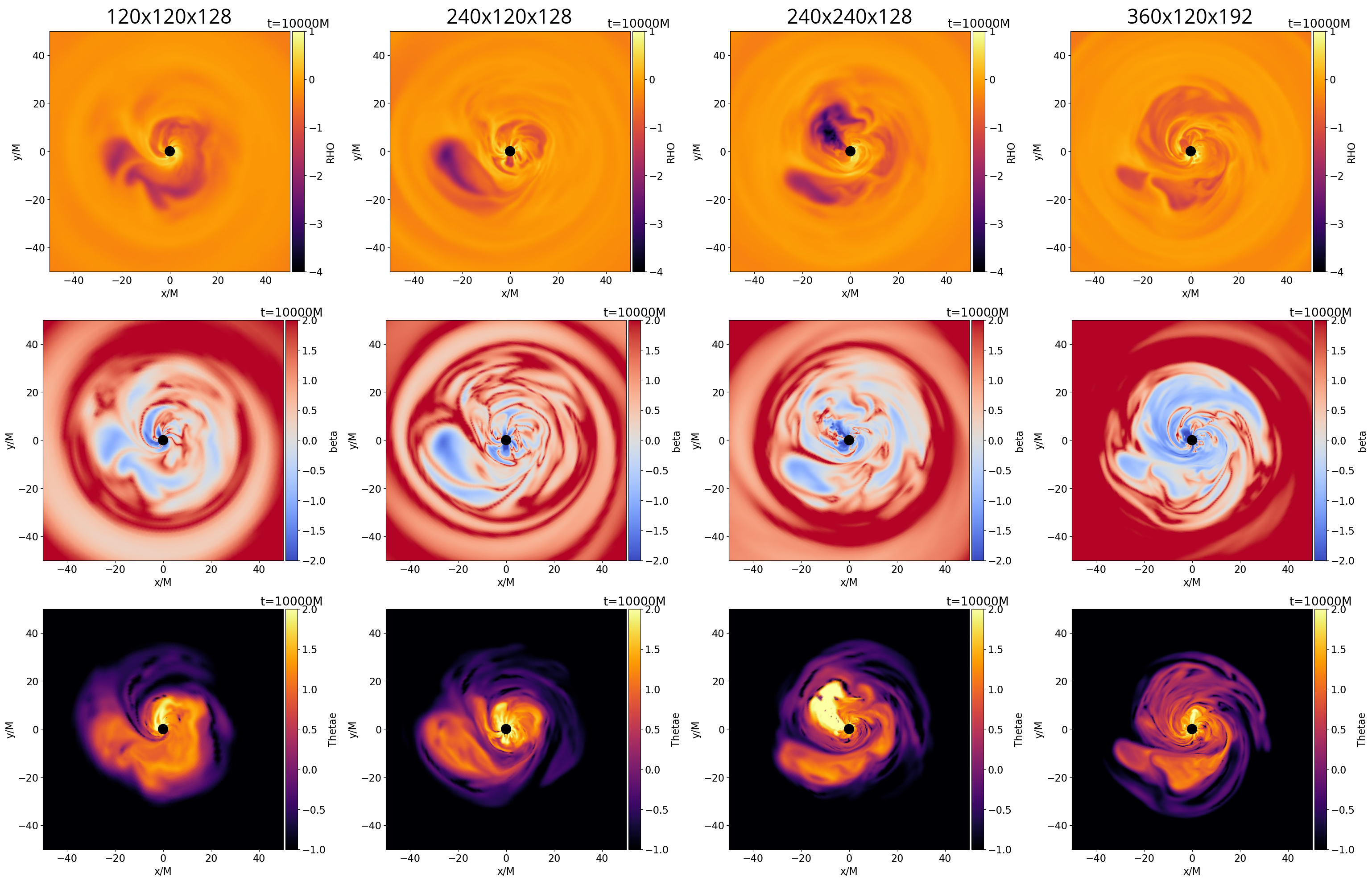} \\
  \caption{Equatorial density (top panels), plasma $\beta$ (middle panels), and K model electron temperature (bottom panels) maps in various-resolution two-temperature MAD
    $a_* = 0$ models at $t = 10,000$M.}\label{fig:res_grmhd}
\end{figure*}

\begin{figure*}
  \centering
  \includegraphics[width=1\linewidth]{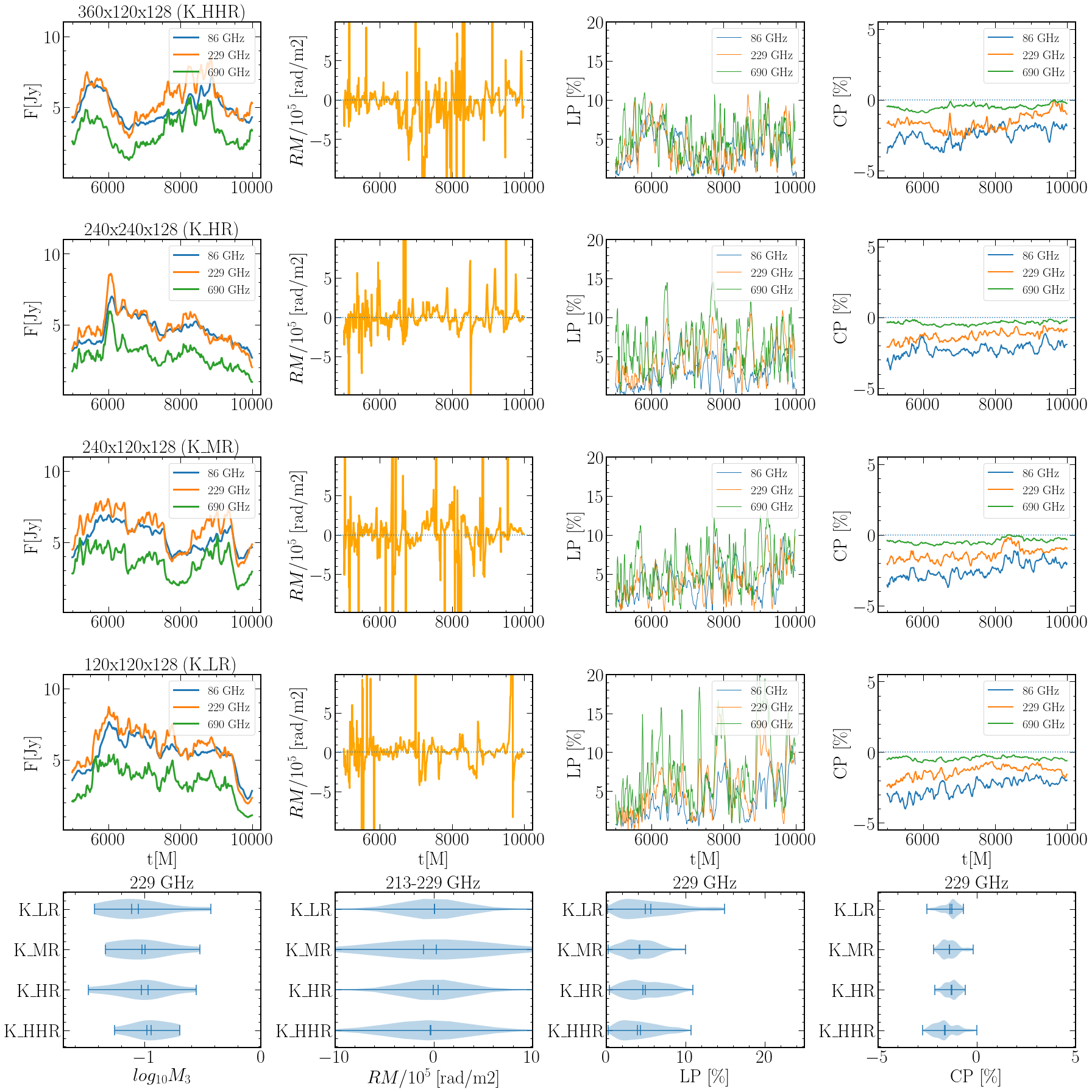}
  \caption{Comparison of Stokes ${\mathcal I}$, RM, LP, and CP
    in two-temperature MAD $a_*=0$ models with increasing grid resolution (Table~\ref{tab:all}). The vertical bars in the violin plots mark mean, median, and upper/lower limits of the distribution.}\label{fig:res_lc}
\end{figure*}

Table~\ref{tab:all} lists four test two-temperature MAD models $a_* = 0$ carried out at different numerical resolutions. 
Figure~\ref{fig:res_grmhd} shows the equatorial cuts through the four models
at the final integration time, $t_f = 10,000$M. 
Figure~\ref{fig:res_lc} displays the corresponding lightcurves for Stokes ${\mathcal I}$, RM, LP,
and CP calculated for the time interval between 5000M and 10,000M at three
observing frequencies assuming the K model for electron temperatures.
In Figure~\ref{fig:res_lc}, the bottom panels display a comparison of the $M_3$,
RM, LP, CP distributions at 229 GHz as a function of grid resolution. There is a very good
agreement between the $M_3$, RM, LP and CP distributions. Although not
shown, this is also true for two other neighboring frequencies of 86 and 690 GHz
at which the modeled source could be observed.  

We conclude that the results
presented in the remainder of this work do not strongly depend on the chosen GRMHD
grid resolution. Around the current level of resolution, which is typically
used for building EHT GRMHD model libraries, all the modeled quantities are
rather settled, and the small variations in observables are due to different
realizations of turbulence in individual runs\footnote{Turbulence is always
initialized by a random perturbation of the gas internal energy.}. We note that we cannot exclude that our models are located at the resolution plateau and that increasing resolution 10 times in each direction may result in the simulation better resolving more physical effects (e.g., better resolution of shocks and reconnection layers may lead to different electron heating in different regions of the flow) that could lead to quantitative and qualitative changes in the emission properties.

\section{Results}\label{sec:results}

\subsection{Turbulent heating vs. ($R~(\beta)$) model}\label{results:K_Rbeta}

\begin{figure*}
  \centering
  \includegraphics[width=1.\linewidth]{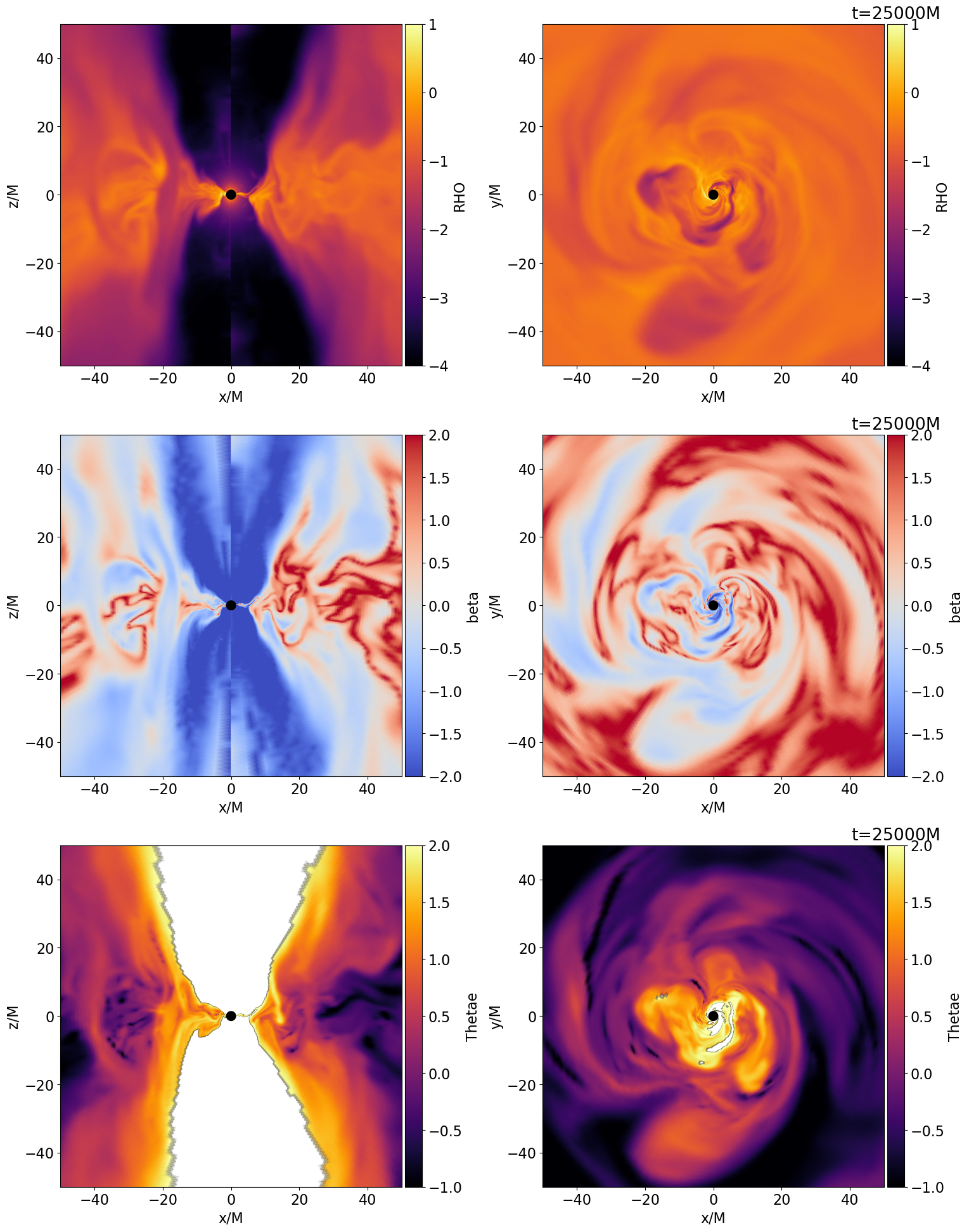}
  \caption{Examples of meridional and equatorial slices showing plasma density (top
    panels), plasma $\beta$ parameter (middle
    panels) and K model electron temperatures (bottom panels) in the fiducial MAD
    $a_* = 0.9375$ simulation at $t = 25,000$M (see Table~\ref{tab:all}). The
    panel with electron temperature masks the uncertain region not taken into account in radiative transfer calculations.}\label{fig:grmhd}
\end{figure*}

The parameters of all the fiducial runs are summarized in Table~\ref{tab:all}. Figure~\ref{fig:grmhd} shows
an example of snapshots from our fiducial two-temperature MAD $a_* = 0.9375$ simulation. The fiducial lightcurves produced by the GRMHD simulations with different parameters are presented in Appendix~\ref{app:lc} and here we discuss distributions of quantities calculated based on these lightcurves.

\begin{figure*}
\centering
\includegraphics[width=1.0\linewidth]{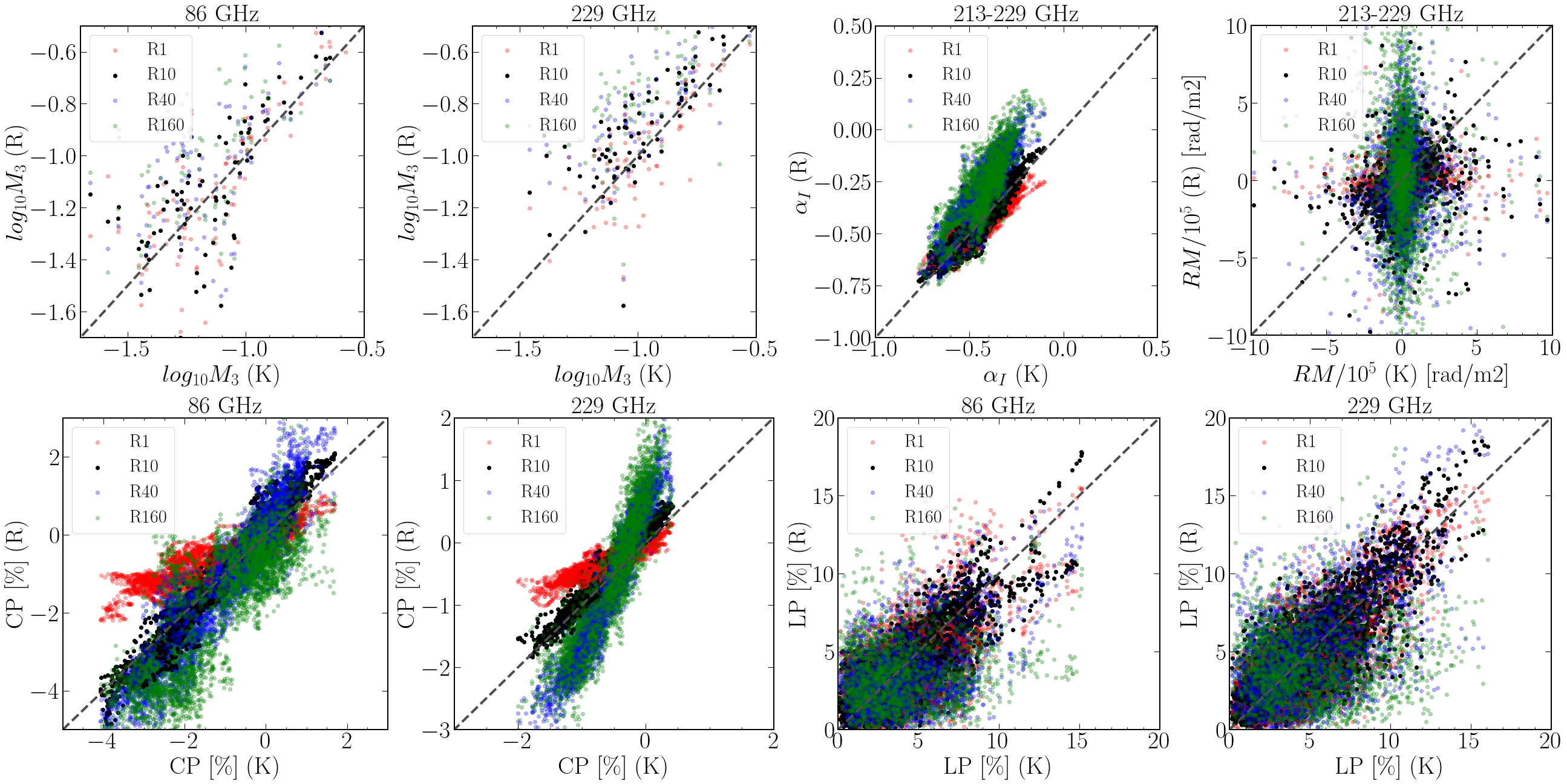}
\caption{Comparison of the net radiative properties for MAD K and R1 - 160
  (spins $a_* = 0, 0.5, 0.9375$ combined)
  models calculated for the default
  viewing angle $i = 160^\circ$.  Top panels show modulation indices of Stokes
  ${\mathcal I}$ light curves at 86 and 229 GHz and spectral slopes and RM
  distributions around 229 GHz.
  In the bottom panels, we show distribution of fractional CP and LP polarizations for 86 and 229 GHz. All net quantities are built based on
  light curves from 15,000 to 30,000M.}\label{fig:hist_MAD}
\end{figure*}

Figure~\ref{fig:hist_MAD} displays the correlations of the observables in
models R1 - 160 and K for the default viewing angle $i=160^\circ$, combined for
black hole spins $a_*=0, 0.5, 0.9375$. In the $M_3$ panels, the K models are
moderately correlated with the R1 - 160 models, and the K models are generally
less variable compared to models R1 - 160. The spectral slope $\alpha_I$ and
CP in model K are highly correlated with those of model R10. This is not the
case for RM and LP, for which the correlation is much weaker.

Figure~\ref{fig:hist_MAD2} shows the comparison of the distributions of the
observables for the models shown in Figure~\ref{fig:hist_MAD}. The averages of
$M_3$, LP, and CP (and standard deviations) in model K are weakly
distinguishable from those of models R1 - 160.

It is useful to look at the properties of the maps used to synthesize the integrated observables. 
Figure~\ref{fig:resolved_pol} shows examples of EHT-like images at 229 GHz
produced by a random snapshot of the $a_* = 0.5$ simulation with R1, R10, R40,
R160 and K electron models for the fiducial viewing angle of $i=160^\circ$. Following \citealt{EHTC:2024VIII} we characterize these images using the following metrics. Net LP and CP are defined as
\begin{equation}
m_{\rm net}\equiv \frac{\sqrt{(\sum_i Q_i)^2 + (\sum_i U_i)^2}}{\sum_i
I}, \,\,\\
v_{\rm net}\equiv \frac{\sum_i V_i}{ \sum_i I_i}.
\end{equation}
The image-averaged LP and CP are defined as
\begin{equation}
m_{\rm avg}\equiv \frac{\sum_i \sqrt{Q_i^2+U_i^2}}{\sum_i I_i},\,\,\\
v_{\rm avg}\equiv \frac{\sum_i |V_i/I_i| I_i}{\sum_i I_i};
\end{equation}
where sums are carried
out over all image pixels. The geometry of LP is described with amplitudes and phases of a complex function, 
\begin{equation}
\beta_{\rm m} \equiv \frac{\int_0^\infty
\int_0^{2\pi} P(\rho,\phi) e^{-i m \phi} \rho d \phi d\rho}{\int_0^\infty \int_0^{2\pi} I(\rho,\phi) \rho d \phi d\rho}
\end{equation}
where
\begin{equation}
P(\rho,\phi)\equiv Q(\rho,\phi)+iU(\rho,\phi)
\end{equation}
is the complex polarization vector and $\rho, \phi$ are the
polar coordinates in the image plane (see
\citealt{palumbo:2020} or \citealt{EHTC:2024VIII} for details). All these metrics are shown in the example snapshot in Figure~\ref{fig:resolved_pol}. 
Here, it is more evident that the resolved images of model K are most similar to model R10 in both LP and
CP. Model R1 differs from K in CP maps. Models R40 - 160 already have a distinct
appearance compared to model K in both LP and CP, as also evident from
the comparison of integrated CP. When scored against existing or future (total
intensity or polarimetric) EHT data, model
K may favor similar accretion flow and geometrical parameters as model R10.

It is important to show that the \trat in the K models does not precisely follow the $R~(\beta)$ model R10, regardless of the fact that \qratio is strongly dependent on $\beta$. An example of \trat maps for a single snapshot of the MAD model with high spin is shown in Figure~\ref{fig:tpte} (top and middle panels). The discrepancy between \trat in models R10 and K indicates that there is no simple relationship between \qratio and \trat as already shown by \citet{ressler:2015} for non-MAD models.
Hence, the exact electron temperature consistency is not the main reason behind the similarity of K and R10 models. 
The difference in \trat between the models is the smallest close to the black
hole, and it increases with radius. The emitting regions close to the black
hole have on average \trat=10. The R10 and K models therefore share the scaling
unit ${\mathcal M}$ to reproduce the same flux, meaning that $\rho$ and $B$ in
the synchrotron radiative transfer coefficients are identical, leading to
similar images. The differences between the model's emissions reflect
differences in $\Theta_e$ (shown in the bottom panels of Figure~\ref{fig:tpte}) and seem small from the observational point of view. Due to the Faraday rotation effects, LP and RM are the most sensitive among our observables to electron temperatures at small and larger radii. This explains the weaker correlation of LP and RM between the R10 and K models visible in Figure~\ref{fig:hist_MAD}. 

\subsection{Importance of other physical effects}

\subsubsection{Effects of radiative cooling}\label{section:cool}

We perform a GRRMHD simulation of the MAD $a_* = 0$ model to
check the impact of radiative cooling on electron temperatures and emission
properties. The radiative
processes included in the GRRMHD simulation include synchrotron emission, self-absorption, Compton scatterings,
and bremsstrahlung. To reduce the large computational cost of models with activated multifrequency
radiative transfer, the interactions between the photon field and the gas are allowed
only within $100$M. Any radiative transfer effects outside of this sphere are
negligible.
When producing emission, the code takes into account the fact that
the combined four-momentum of gas and radiation has to be
conserved. Emission/absorption/scattering can therefore exert force on the gas and
change the electron thermodynamics (radiative cooling or heating). In the
models considered here, the average optical thickness of the gas is $\tau \ll 1$
and luminosities are strongly sub-Eddington, $L \ll L_{\rm Edd}$; hence, it
is expected that in models with activated radiative transfer, mainly electron
thermodynamics is affected and mainly due to synchrotron cooling. Other
effects are weak for the accretion rates considered.
In Appendix~\ref{app:grrmhd}, we present quality factors for the GRRMHD
simulation to demonstrate that the interactions between gas and radiation are
resolved by an adequate number of Monte Carlo samples. 

Figure~\ref{fig:GRRMHD} displays volume-averaged radial profiles of density,
magnetic field, \trat (or $R$), and $\Theta_e$ in the GRMHD and GRRMHD model with the K
heating mechanism and ${\mathcal M}=5\times10^{11}$ at different time moments, earlier and later in the evolution. All quantities are
additionally weighed by the $\rho B$ factor to show quantities averaged over
the regions where most of the synchrotron emission comes from (the $\rho B$ factor is a major factor in synchrotron emissivity; see, e.g. \citealt{pandya:2016} and their Eq.~24 or Eq.~29). There is a small difference in the radial profiles
$\Theta_e$ and $R$ in models with and without radiative
effects, and the difference is due to the different realization of turbulence in
each model. Unsurprisingly, at later times in both models, $R \sim 10 $ for
$r<20$M, consistent with the results in Section~\ref{results:K_Rbeta}.
Figure~\ref{fig:cooling} shows that there is no difference in radiative
characteristics between the GRMHD (K) and GRRMHD (Kcool) models based on
$15,000-30,000$M light curves.
The total 230~GHz flux of model K averaged over the above time period is 2.99
and 2.93 Jy in models with and without radiative effects, respectively. This
suggests that the radiating models simply do not cool fast enough
to significantly affect the electron temperatures. This is expected
considering very small accretion rates onto Sgr~A*.

Since the test model with cooling has the highest ${\mathcal M}$
among all K models considered in this work (see Table~\ref{tab:all}), all other K models shown in the paper are expected to cool down even slower than the test case. Radiative cooling effects in MADs with turbulent heating may be neglected when modeling Sgr~A* millimeter emission. 

%We have also calculated long-duration light-curves based on one of the GRRMHD
%simulation (shown in Figure~\ref{fig:lc_MADa0}, bottom panels labeled
%Kcool). The statistical properties of the light-curves with cooling effects
%are the same compared to non-radiating models.

\subsubsection{Effects of nonthermal electrons}

Figure~\ref{fig:hist_MAD_comparison_to_obs_nonthermal} shows the comparison of the
observational characteristics of thermal and nonthermal K fiducial
runs. Non-thermal models are slightly less variable compared to the thermal
models. The spectral slope $\alpha_I$ and the RM and the LP of both thermal and nonthermal
models are strongly correlated (the correlation is much stronger compared to the $R~(\beta)$ models). It is evident that the total intensity and
LP are not sensitive to details of the distribution function
for the considered part of the electromagnetic spectrum. CP, on the other hand, is
different in both models; hence it may be used to discriminate between
distribution functions. We note that CP is usually a more permissive
observational constraint compared to LP (see, e.g., \citealt{EHTC:2024VIII}).
Overall, with the exception of CP, all radiative properties of our simple nonthermal models are tightly correlated with their
purely thermal counterparts. 

\subsection{Fiducial models vs. ALMA constraints}\label{sec:comparison}

Having checked how the results depend on numerical resolution effects and the two most relevant physical effects,
we can now compare the thermal two-temperature models directly to observations of plasma around Sgr~A*. 

Figure~\ref{fig:hist_MAD_comparison_to_obs} shows the comparison of the prograde
models K, at various viewing angles, to Sgr~A* observables detected by the
Atacama Large Millimeter/submillimeter Array (ALMA) at
86 and 213 - 229 GHz in 2017, April
\citep{wielgus:2022a,wielgus:2022b,wielgus:2024}. The modulation index at both
frequencies increases with black hole spin and inclination angle. In
general, all models are too variable to match the observed variability (this is
also the case for the R1 - 160 models found in \citealt{EHTC:2022V} and shown here in Figure~\ref{fig:hist_MAD2}), but the data
strongly favor lower spins at lower inclination angles, for which $M_3$ is the
smallest. All models are too optically thin when compared to observations (as
is also the case for the R1 - 160 models; compare our
Figure~\ref{fig:hist_MAD2} with
Figure~\ref{fig:hist_MAD_comparison_to_obs_retro}, and see also \citealt{ricarte:2023}).  All models also
require an external Faraday screen to explain the observations of RM, but we
notice that models with low viewing angles ($i=160^\circ$, $150^\circ$) best recover
the RM standard deviation toward the source. LP of the
$i = 160^\circ - 150^\circ$ models is consistent with both 229 and 86 GHz
measurements. Models with $a_* = 0$ recover CP at 229 GHz but fail to recover
CP at 86 GHz. On the other hand, models with $a_* = 0.9375$ fail to recover CP
at 229 GHz, but do recover CP at 86 GHz. Only the model with $a_* = 0.5$ does
reasonably well with regard to CP at both frequencies. It is possible that a
model with a black hole spin in between 0 and 0.5 would be a better fit to the ALMA
CP data. Among prograde MADs, the model with $a_* = 0.5$ is the closest to the
ALMA observations overall.

Although turbulence in the retrograde models reached a quasi-stationary state
within the emission zone, they are reported separately because they are
integrated over shorter times compared to the prograde
models. Figure~\ref{fig:hist_MAD_comparison_to_obs_retro} compares the statistical
properties of the light curves produced by retrograde MADs. The two-temperature
MAD with $a_* = -0.5$ performs better compared to MAD $a_* = -0.9375$ as they are
less variable, but worse than MAD $a_* = 0.5$, because prograde models better
recover CP at both frequencies.

\section{Discussions}\label{sec:discussion}

In this work, we have carried out several two-temperature MAD simulations with
the main purpose of investigating the impact of various numerical parameters
on the emission properties of these models, in particular grid resolution, but
we also investigate the impact of physical effects. 

Regarding numerical parameters, the two-temperature MAD simulations with
turbulent heating of electrons are well converged with respect to the GRMHD
grid resolution in agreement with the only published emission convergence test based
on non-MAD models \citep{ressler:2017}.
%Fast-light simulations are also extremely good approximation when computing (sub-)millimeter emission from the time-varying two-temperature MAD models.  

Regarding physical parameters, radiative cooling can be safely omitted in
Sgr~A* two-temperature MAD models with turbulent heating, as radiative effects
do not significantly affect the electron temperature evolution. These effects
may be important when modeling M87, which is a brighter source. We have also
carried out the first exploratory survey of models in which turbulent heating
is dissipated into a nonthermal electron distribution function. Our nonthermal
models display very similar characteristics compared to the thermal
models. However, including the physics of nonthermal electrons in the emission models slightly decreases the model variability and alters the CP compared to purely thermal models. Other, also spatially variable parameters of the $\kappa$ distribution function could be considered and change our result \citep[e.g.,][]{davelaar:2019}.

Long-duration two-temperature simulations indicate that the thermal models of K are
only roughly approximated with the R10 models. Using total intensity observables, \citet{Mizuno:2021} found that the best match to turbulent heating models are R5 models. Given that their electron temperatures are defined differently (compare their Eq.~8 with our Eq.~\ref{eq:thetae}) and are a factor of $\sim2$ smaller than ours, the best match with R10 found here is consistent with the previous finding.

In Section~\ref{sec:comparison} we show that
none of our long-duration two-temperature K simulations recover the general
properties of light curves of Sgr~A* observed by ALMA ideally, but the ALMA
data favor models with lower viewing angles $i = 150^\circ - 160^\circ$ for
which the variabilities of Stokes ${\mathcal I}$ and RM are the smallest. All
models have great difficulty recovering the observed model spectral index, in
excellent agreement with the previous two-temperature study by \citealt{dexter:2020b}. All models recover the observed net LP but require the introduction of an additional external Faraday screen to recover the observed RM. 
Only the $a_* = 0.5$ model recovers Sgr~A* net CP at two observed frequencies
simultaneously, and it is possible that a spin between 0 and 0.5 would produce even better results. 
The preferred moderate/low spin value is consistent with spin estimates based on
two-temperature models of \citealt{dexter:2020b}, although for different than
the turbulent heating scenario, different initial conditions, and different observational datasets.
Here we also study the evolution of two-temperature MADs for low and high
spins for a significantly longer time compared to \citealt{dexter:2020b}. 

It is possible that ALMA observes varying compact emission together with less
varying extended emission that is not captured in our simulations (C.F. Gammie
2025, private communication); hence, it is preferable to compare models with
EHT images (or best with the future EHT movies) rather than integrated
quantities. Compared to EHT data, the K models are expected to favor a
parameter space similar to the R10 models. Polarimetric EHT data of Sgr~A*
have not yet ruled out all MAD R10 models. In fact, 10 out of 20 MAD models
passing are R10 models (see Fig.~8 in \citealt{EHTC:2024VIII}). Based on the
similarity between the appearance of R10 and K, we argue that the turbulent
electron heating scenario could still be consistent with the observations of
Sgr~A*. It is interesting to note that in \citealt{EHTC:2024VIII}, MAD model
R10 is favored for a set of intermediate viewing angles ($i = 150^\circ - 110^\circ$) and positive and negative spin values. K models could narrow the range of these two geometric parameters. 

Based on joint total intensity and polarimetric EHT data, the current best-bet MAD model for Sgr~A* is R160, not R10 \citep{EHTC:2024VIII}. We should discuss the present results in the context of other models for \qratio.
As shown by \citet{dexter:2020b} assuming the \citet{howes:2010} turbulent heating
model for \qratio, there is a negligible difference in the MAD electron
temperatures compared to the \citet{kawazura:2019} model. The reconnection \qratio
model of \citealt{rowan:2017}, on the other hand, may result in slightly lower
(but also around R10) electron temperatures with weaker plasma-$\beta$
dependency \citep{dexter:2020b,Mizuno:2021}. Lower electron
temperatures could make the MAD models optically thicker and less varying \citep{chan:2024},
thus pushing them toward the right direction, but this remains to be thoroughly examined.
Weaker dependence on $\beta$ in models with
reconnection heating can result in more disklike emission of MADs than disk/jetlike emission \citep{chael:2018}, making it more difficult to
explain the nearly flat radio spectrum of Sgr~A*
\citep{moscibrodzka:2013}. The radio spectral slope dependency on \qratio should be carefully checked with high-resolution (jet-resolving) two-temperature MAD models.
However, whichever \qratio scenario above (or a combination of scenarios) is
selected, the two-temperature MAD models will have difficulty in naturally developing R160-like electron temperatures near the event horizon. Assuming that
MAD is present in the Sgr~A* system, this may suggest that $R~(\beta)$ models are
inappropriate for MADs (we briefly discuss prospects on how to resolve this issue in Appendix~\ref{app:rlow}).
%It is also possible that electron distribution functions in MADs are non-thermal but different from the non-thermal models presented in this work, which are nearly indistinguishable from the thermal models. 
Alternatively, comparisons of instantaneous model snapshots with time-averaged EHT images synthesized from the time-varying EHT data sets so far (see \citealt{EHTC:2022V,EHTC:2024VIII}) may be introducing some bias in the inference of the electron temperature parameter.

Our simulations are run with constant adiabatic index $\gamma_{\rm ad} = 13/9$, which is not a
self-consistent assumption in regions where electrons are mildly or
subrelativistic. Recent simulations of \citet{narayan:2022} suggest that the MAD
dynamics is not affected by the choice of $\gamma_{\rm ad}$. It is unclear how
different, or varying, $\gamma_{\rm ad}$ alters the discussed total
intensity and polarimetric radiative properties of two-temperature MADs. This topic should be investigated in the future. 

\begin{figure}
  \centering
\includegraphics[width=1.0\linewidth]{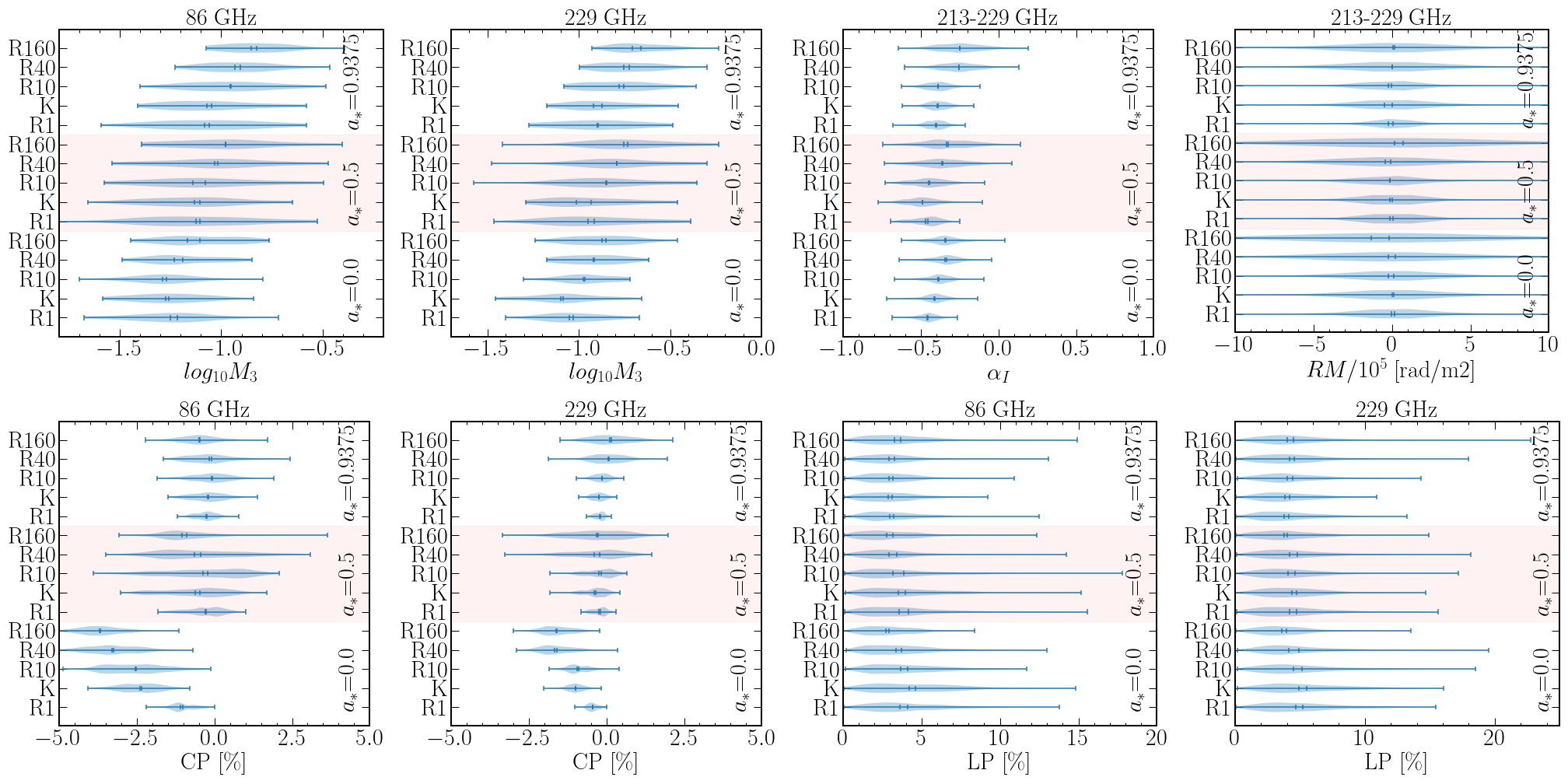}
\caption{Comparison of distributions of various observables in MAD
  models K and R1 - 160 observed at $i = 160^\circ$. The light
  red background is introduced to visually separate models with different
  spins ($a_* = 0, 0.5, 0.9375$). }\label{fig:hist_MAD2}
\end{figure}

\begin{figure}
  \centering
\includegraphics[width=1.0\linewidth]{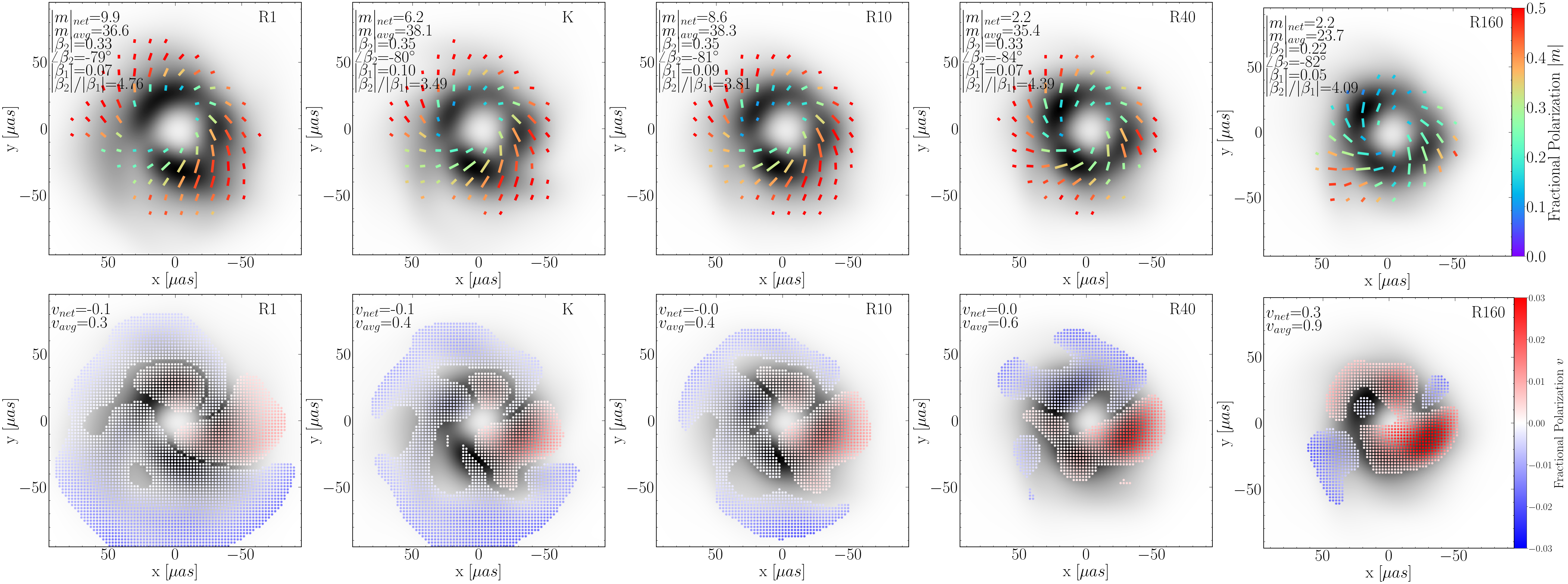}
\caption{Comparison of snapshot images of MAD $a_* = 0.5$ models K, R1, R10, R40 and R160 at
  frequency of 229 GHz for the fiducial viewing angle $i = 160^\circ$. Images are blurred by a Gaussian kernel with
  FWHM = 20 $\mu as$ to imitate the resolution of EHT. In the top panels, the gray scale is used for total
  intensity, and color ticks illustrate the LP of the
  emission. The linearly polarized fraction of total intensity is marked with color, and tick length
  is proportional to $P \equiv \sqrt{Q^2+U^2}$. We do not show LP in
  regions where flux in $P$ drops below 20\% of its maximum. In the bottom
  panels, the fractional CP
  is plotted as points. We do not show CP in regions where $|V|$ drops below 5\% of its
  maximum. K models are close to R10 models, but the details of the polarimetric
  characteristics are slightly different.}\label{fig:resolved_pol}
\end{figure}

\begin{figure}
  \centering
\includegraphics[width=1\linewidth]{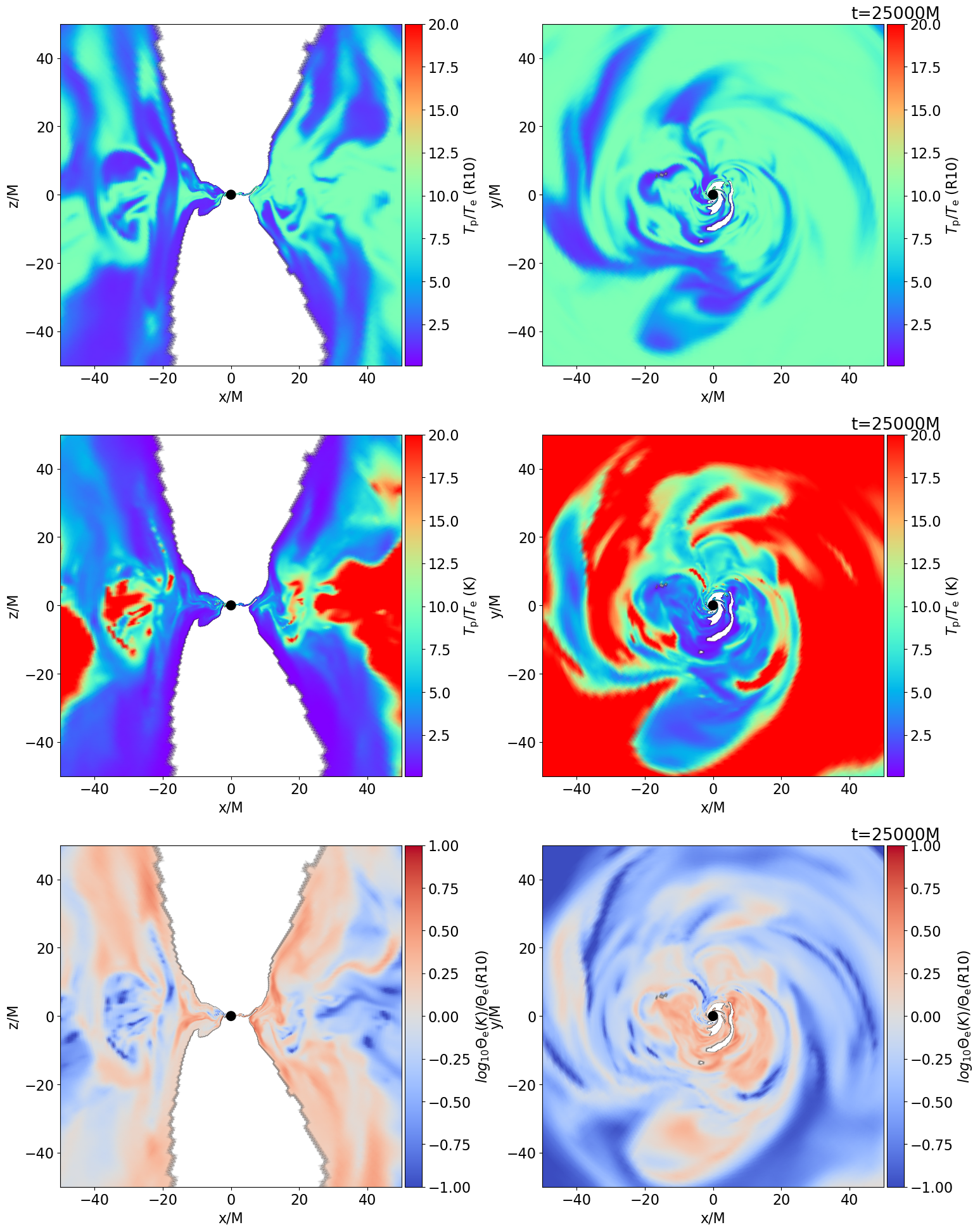}\\
\caption{Examples of meridional and equatorial slices showing \trat (top
  panels show \trat in model R10 and middle panels show \trat in model K) and
  the ratio of K and R10 $\Theta_e$ (bottom panels) in MAD $a_* = 0.9375$ at $t = 25,000$M.  The uncertain regions, not taken into account in radiative transfer calculations, are masked.}\label{fig:tpte}
\end{figure}

\begin{figure*}
\centering
\includegraphics[width=1\linewidth]{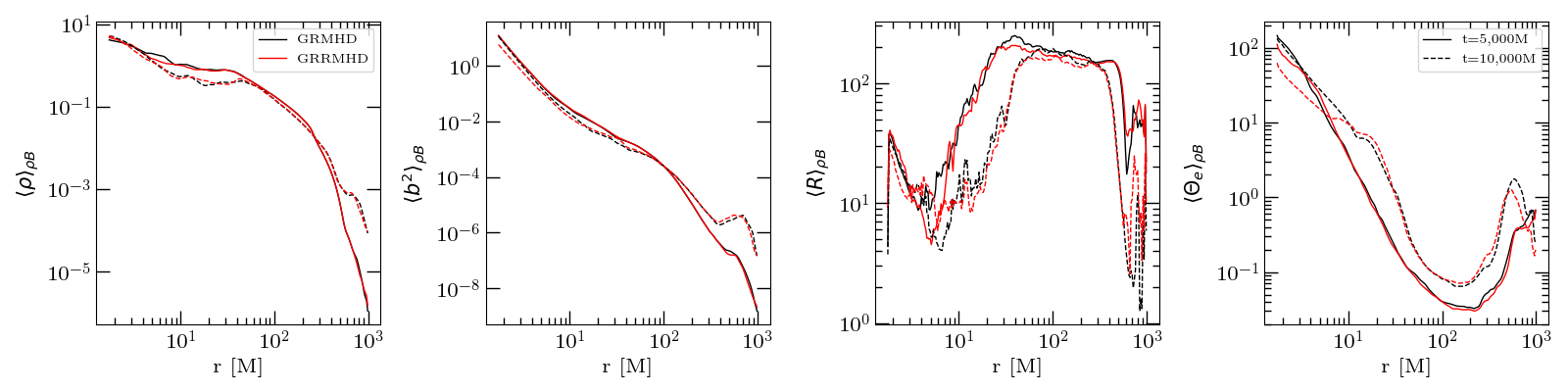}
\caption{Comparison of radial profiles of density, magnetic field strength
  (square), proton-to-electron temperature ratio ($R$) and electron temperatures in
  GRMHD (black lines) and GRRMHD (red lines) models. All quantities are dimensionless or in code units,
  and they are all are angle-averaged, where averaging of a quantity $Q$ is
  defined as $\left< Q(r,t) \right> \equiv \int_0^{2\pi} \int_0^{\pi}
  Q(r,\theta,\phi,t) \sqrt{-g} d\theta d\phi / \int_0^{2\pi} \int_0^{\pi}
  \sqrt{-g} d\theta d\phi$ where $g$ is the metric determinant. All quantities
  are also weighted with the $\rho B$ factor to show quantities averaged over regions where most of the synchrotron emission comes from. Different line styles correspond to different time moments of the MAD evolution.}\label{fig:GRRMHD}
\end{figure*}

\begin{figure*}
\centering
\includegraphics[width=1\linewidth]{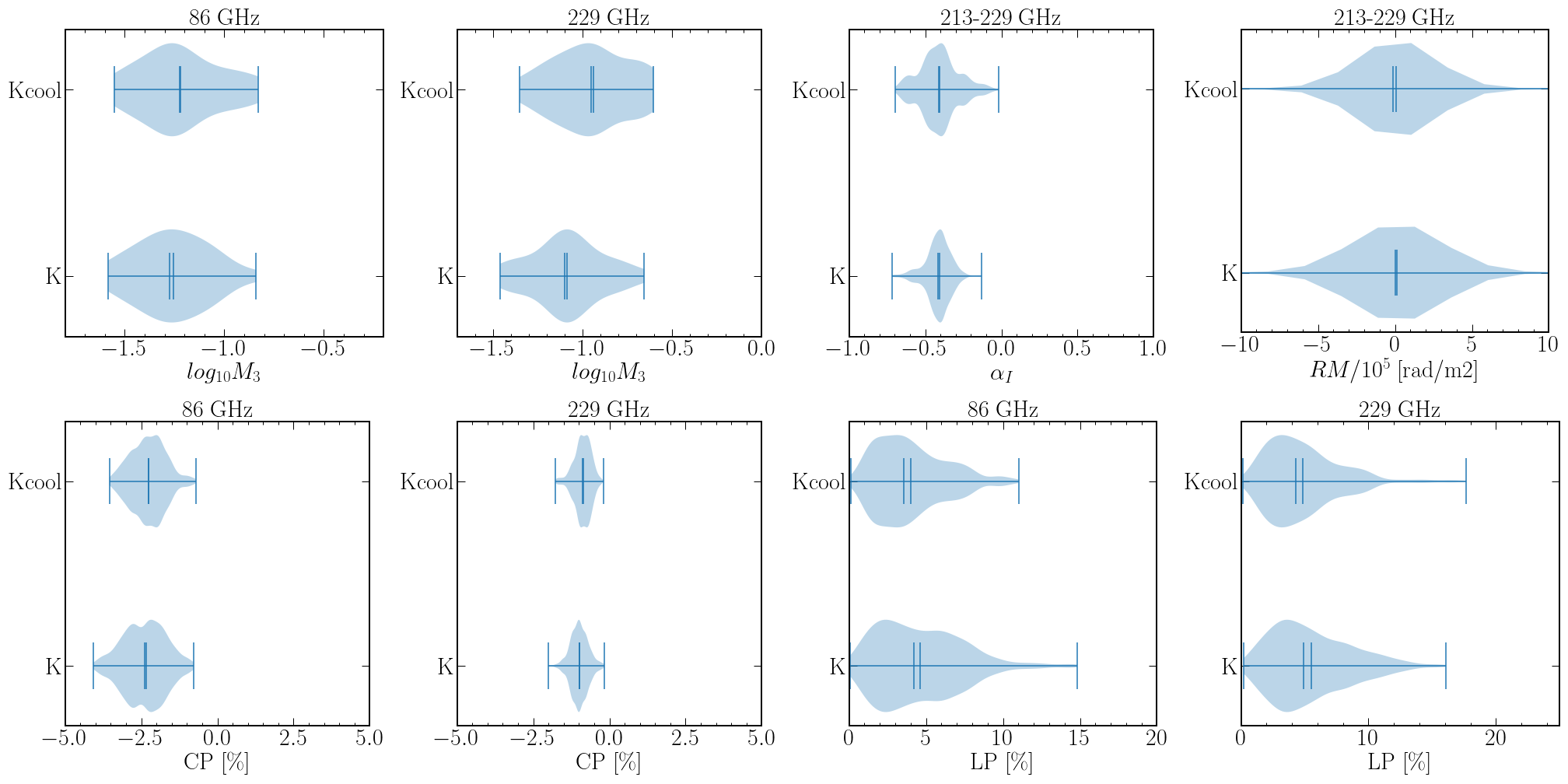}
\caption{Comparison of radiative characteristics of the GRMHD (K) and GRRMHD (Kcool) models
  with $a_* = 0$ spin computed for light curves in the time interval $15,000-30,000$M.}\label{fig:cooling}
\end{figure*}

\begin{figure*}
  \centering
\includegraphics[width=1\linewidth]{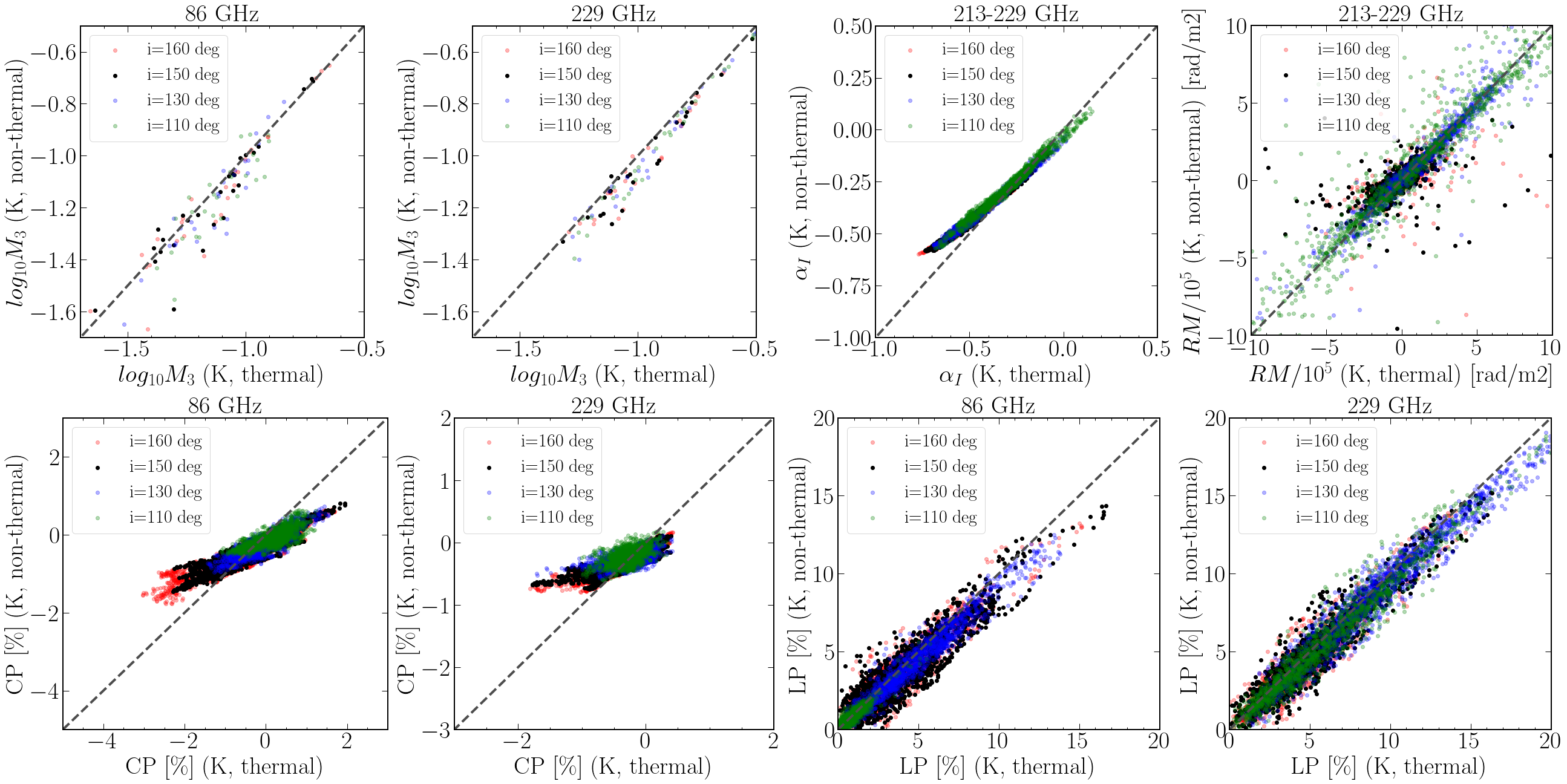}
\caption{Comparison of the radiative characteristics of thermal and
    nonthermal MAD models with $a_* = 0.5$. All panels are the same as in
    Figure~\ref{fig:hist_MAD} but the colors indicate models with different
    viewing angles rather than different $R_{\rm high}$ parameters.}\label{fig:hist_MAD_comparison_to_obs_nonthermal}
\end{figure*}

\begin{figure*}
  \centering
  \includegraphics[width=1\linewidth]{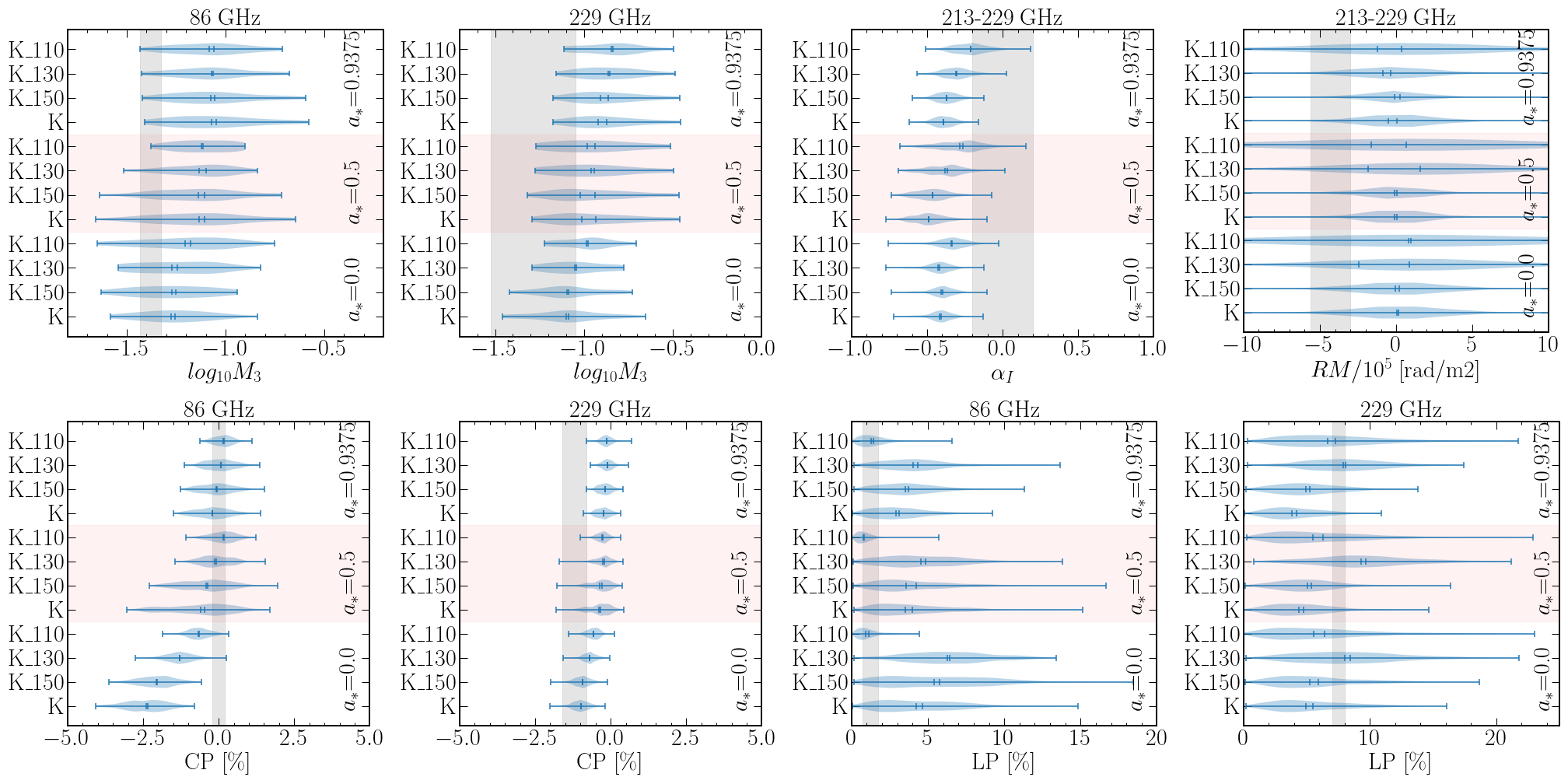}
\caption{Comparison of radiative characteristics of MAD models K
  ($a_* = 0, 0.5, 0.9375$) at a default viewing angle of $i = 160^\circ$ (K) and
  three additional angles $i = 150^\circ, 130^\circ, 110^\circ$ deg (K\_150, K\_130, K\_110) to observations of Sgr~A*
  collected by ALMA in April 2017 (86 GHz data collected on April 3rd and 230 GHz data collected on April 6-11) and presented in
  \citet{wielgus:2022a,wielgus:2022b,wielgus:2024} (gray bands).  The light
  red background is introduced to visually separate models with different
  spins.}\label{fig:hist_MAD_comparison_to_obs}
\end{figure*}

\begin{figure*}
  \centering
\includegraphics[width=1.0\linewidth]{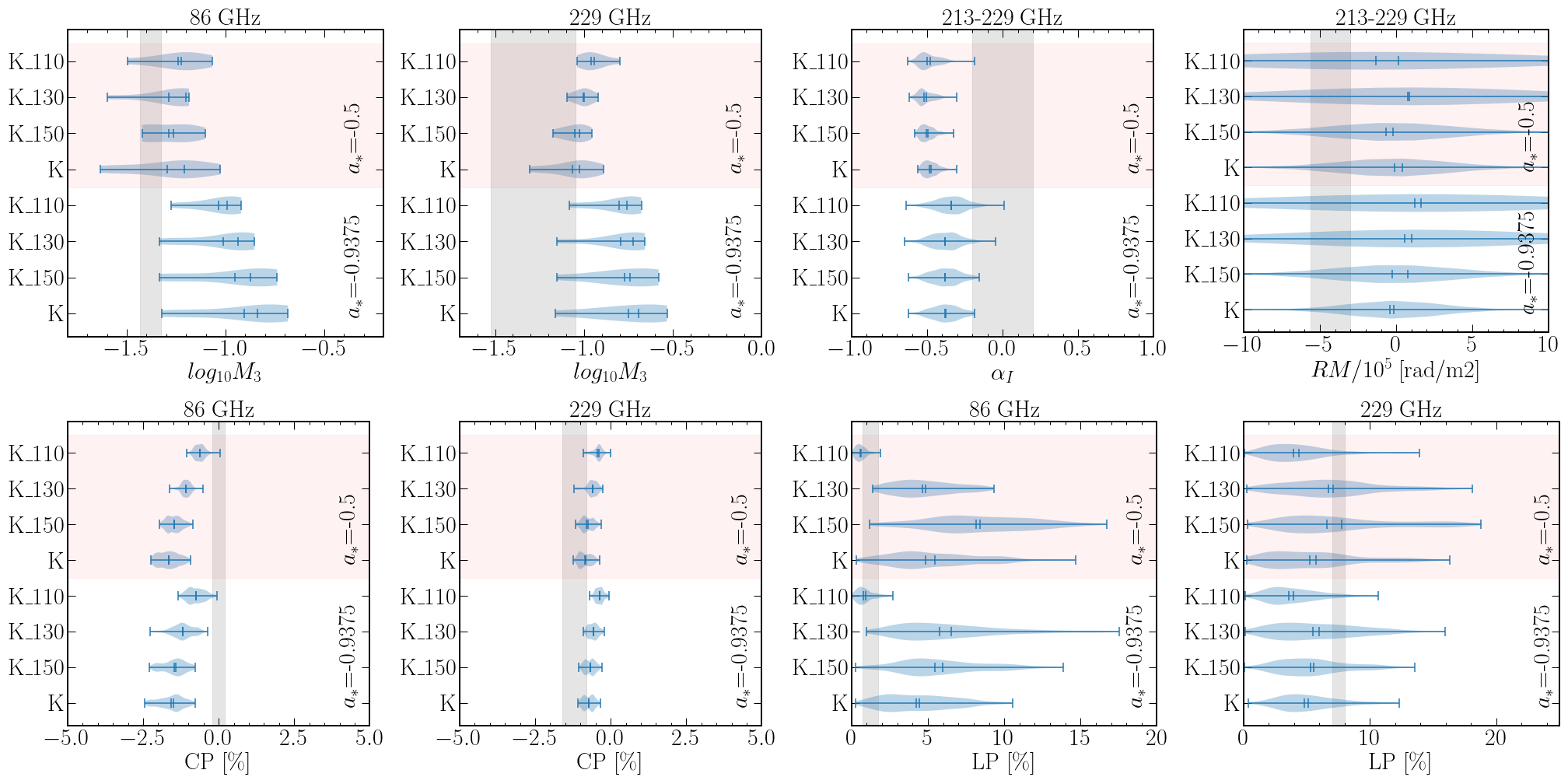}
\caption{Same as Figure~\ref{fig:hist_MAD_comparison_to_obs} but for
  retrograde models K.}\label{fig:hist_MAD_comparison_to_obs_retro}
\end{figure*}

The recent manuscript by \citet{salas:2024}, which was posted after this paper
was submitted, models Sgr~A* variability in total intensity with
two-temperature $a_* = 0.94$ MAD models combining reconnection heating, varying $\gamma_{\rm ad}$ index, and radiative cooling. Although we agree with their statement that the physical two-temperature models are typically less variable compared to the $R~(\beta)$ models (via $M_3$ characteristics; see our Figure~\ref{fig:hist_MAD2}, second top panel), further detailed comparisons of our results to \citet{salas:2024} are difficult due to different assumptions in electron heating model, equation of state, and radiative cooling methodology. Interestingly, \citet{salas:2024} models, as ours, still do not produce variability fully consistent with the observational data of \citet{wielgus:2022a}. 

The ideal GRMHD models considered here do not include many other potentially important effects. 
Additional physics, such as resistivity \citep[e.g.,][]{Vos:2024}, viscosity, plasma composition \citep{emami:2021,wong:2022}, anisotropic pressures \citep[e.g.,][]{foucart:2017}, or anisotropic electron distribution functions \citep[e.g.,][]{galishnikova:2023}, may all impact the radiative characteristics of MADs. 
The study of the influence of all these effects on observed quantities goes beyond the scope of the present work. The current simulation set constitutes a reference point for future investigations.

\begin{acknowledgements}
  The authors thank Maciek Wielgus for providing ALMA data and comments on the paper.
  We also thank Andrew Chael, Charles Gammie, and Ben Ryan for many of their comments. We gratefully acknowledge the HPC RIVR
consortium (www.hpc-rivr.si) and EuroHPC JU (eurohpc-ju.europa.eu) for funding
this research by providing the computing resources of the HPC system Vega at the
Institute of Information Science (www.izum.si).
\end{acknowledgements}

{\it Software} \texttt{ebhlight} \citep{ryan:2015,ryan:2017,ryan:2018}, \texttt{ipole} \citep{moscibrodzka:2018}, \texttt{python} \citep{oliphant:2007}, \texttt{matplotlib} \citep{hunter:2007}.

%\bibliography{local}{}
%\bibliographystyle{aa}

\begin{table*}
\centering
\begin{center}
  \begin{tabular}{cccccccccccc}
  \hline
   $a_*$ &  ID & $R_{\rm low}$ & $R_{\rm high}$ & i & $\frac{\mathcal M}{10^{17}}$ & $\left<\dot{M}\right>$&$t_{\rm
    s}$& $t_{\rm f}$ &$\Delta t$& Resolution & Cooling \\
         &     &               &                & (deg) & & ($M_\odot/yr$) &
  (M) &(M) & (M) & & \\
  \hline
  \multicolumn{11}{c}{Convergence Study Models} \\
  \hline
   0& K\_HHR& -&-& 160 & 5 & $1.0 \times 10^{-8}$ & 5000 & 10,000 & 10 & $360 \times 120 \times 192$ & no\\
   0& K\_HR & -&-& 160 & 5 & $1.1 \times 10^{-8}$ & 5000 & 10,000 & 10 & $240 \times 240 \times 128$ & no\\
   0& K\_MR & -&-& 160 & 5 & $1.0 \times 10^{-8}$ & 5000 & 10,000 & 10 & $240 \times 120 \times 128$ & no\\
   0& K\_LR & -&-& 160 & 5 & $1.2 \times 10^{-8}$ & 5000 & 10,000 & 10 & $120 \times 120 \times 128$ & no\\
  \hline
  \multicolumn{11}{c}{Fiducial Models} \\
  \hline
%   -0.9375 &K & - & -  & 160,150,130,110 & 2,2,2,2 & & 10,000 & 14,000 & 10 & 266x120x128& no\\
 -0.9375 &K & - & -  & 160 & 2 & $4.6 \times 10^{-9}$ & 10,000 & 14,000 & 10 &$266 \times 120 \times 128$& No\\ 
 -0.9375 &K & - & -  & 150& 2 & $4.6 \times 10^{-9}$& 10,000 & 14,000 & 10 & $266 \times 120 \times 128$& No\\
 -0.9375 &K & - & -  & 130 & 2 &$4.6 \times 10^{-9}$ & 10,000 & 14,000 & 10 &$266 \times 120 \times 128$& No\\
  -0.9375 &K & - & -  & 110 & 2 & $4.6 \times 10^{-9}$& 10,000 & 14,000 & 10 &$266 \times 120 \times 128$& No\\
%   -0.5 &K & - & -  & 160,150,130,110 & 5,4.5,4,4 & &10,000 & 14,000 & 10 & 240x120x128& no\\
-0.5 &K & - & -  & 160 & 5 & $6.9 \times 10^{-9}$ &10,000 & 14,000 & 10 & $240 \times 120 \times 128$& No\\
-0.5 &K & - & -  & 150 & 4.5 &$6.2 \times 10^{-9}$ &10,000 & 14,000 & 10 &$240 \times120\times128$& No\\
-0.5 &K & - & -  & 130 & 4& $5.5 \times 10^{-9}$ &10,000 & 14,000 & 10 & $240 \times 120 \times 128$& No\\
-0.5 &K & - & -  & 110 & 4 & $5.5 \times 10^{-9}$&10,000 & 14,000 & 10 & $240 \times 120 \times 128$& No\\
%0 & R1-160 & 1 & 1,10,40,160 & 160&3,5,7,10 & &15,000 & 30,000 & 10 & 240x120x128& no\\
0 & R1 & 1 & 1 & 160&3 & $4.2 \times 10^{-9}$ &15,000 & 30,000 & 10 &
$240 \times 120 \times 128$& No\\
0 & R10 & 1 & 10 & 160&5 & $7 \times 10^{-9}$&15,000 & 30,000 & 10 & $240
\times 120 \times 128$& No\\
0 & R40 & 1 & 40 & 160&7 & $9.8 \times 10^{-9}$&15,000 & 30,000 & 10 & $240
\times 120 \times 128$& No\\
0 & R160 & 1 & 160 & 160&10 &$1.4 \times 10^{-8}$ &15,000 & 30,000 & 10 & $240
\times 120 \times 128$& No\\
%0 & K & - & - & 160,150,130,110 & 5,5,4,4 & &15,000 & 30,000 & 10 & 240x120x128& no\\
0 & K & - & - & 160 & 5 & $7 \times 10^{-9}$ &15,000 & 30,000 & 10 & $240
 \times 120 \times 128$& No\\
0 & K & - & - & 150 & 5 & $7 \times 10^{-9}$ &15,000 & 30,000 & 10 & $240
\times 120 \times 128$& No\\
0 & K & - & - & 130 & 4 &$5.6 \times 10^{-9}$ &15,000 & 30,000 & 10 & $240
\times 120 \times 128$& No\\
0 & K & - & - & 110 & 4 &$5.6 \times 10^{-9}$ &15,000 & 30,000 & 10 & $240
\times 120 \times 128$& No\\
%0.5 &R1-160 & 1 & 1,10,40,160  & 160& 2,3,5,8 & &15,000 & 30,000 & 10 & 240x120x128& no\\
 0.5 &R1-160 & 1 & 1  & 160& 2 & $2.7 \times 10^{-9}$ &15,000 & 30,000 & 10 & $240\times120\times128$& No\\
0.5 &R10 & 1 & 10  & 160& 3 & $4.1 \times 10^{-9}$&15,000 & 30,000 & 10 & $240\times120\times128$& No\\
0.5 &R40 & 1 & 40  & 160& 5 & $6.9 \times 10^{-9}$&15,000 & 30,000 & 10 & $240\times120\times128$& No\\
0.5 &R160 & 1 &160  & 160& 8 & $1.1 \times 10^{-8}$&15,000 & 30,000 & 10 & $240\times120\times128$& No\\
%0.5 &K & - & -  & 160,150,130,110 & 3,3,3,3 & &15,000 & 30,000 & 10 & 240x120x128& no\\
0.5 &K & - & -  & 160 & 3 &$4.1 \times 10^{-9}$ &15,000 & 30,000 & 10 & $240\times120\times128$& No\\
0.5 &K & - & -  & 150 & 3 & $4.1 \times 10^{-9}$&15,000 & 30,000 & 10 & $240\times120\times128$& No\\
0.5 &K & - & -  & 130 & 3 & $4.1 \times 10^{-9}$&15,000 & 30,000 & 10 & $240\times120\times128$& No\\
0.5 &K & - & -  & 110 & 3 & $4.1 \times 10^{-9}$&15,000 & 30,000 & 10 & $240\times120\times128$& No\\
%0.9375&R1-160 & 1 & 1,10,40,160& 160 & 1.5,2,4,6 & &15,000 & 30,000 & 10 &266x120x128& no\\
0.9375&R1 & 1 & 1& 160 & 1.5 & $2.0 \times 10^{-9}$&15,000 & 30,000 & 10 &$266\times120\times128$& No\\
0.9375&R10 & 1 & 10& 160 & 2 & $2.7 \times 10^{-9}$ &15,000 & 30,000 & 10 &$266\times120\times128$& No\\
0.9375&R40 & 1 & 40& 160 & 4 & $5.5 \times 10^{-9}$&15,000 & 30,000 & 10 &$266\times120\times128$& No\\
0.9375&R160 & 1 & 160& 160 & 6 &$8.3 \times 10^{-9}$ &15,000 & 30,000 & 10 &$266\times120\times128$& No\\
%0.9375&K & - & - & 160,150,130,110 & 2,2,2,2 & &15,000 & 30,000 & 10 &266x120x128& no\\
0.9375&K & - & - & 160 & 2 & $2.7 \times 10^{-9}$&15,000 & 30,000 & 10 &$266\times120\times128$& No\\
0.9375&K & - & - & 150 & 2 & $2.7 \times 10^{-9}$&15,000 & 30,000 & 10 &$266\times120\times128$& No\\
0.9375&K & - & - & 130 & 2 & $2.7 \times 10^{-9}$&15,000 & 30,000 & 10 &$266\times120\times128$& No\\
0.9375&K & - & - & 110 & 2 & $2.7 \times 10^{-9}$&15,000 & 30,000 & 10 &$266\times120\times128$& No\\
\hline
   \multicolumn{11}{c}{Exploratory Models} \\
   \hline
   0 & K, cooling & - & - & - &  5 & $8 \times 10^{-9}$  & 15,000 & 30,000 & 10 & $240\times120\times128$ & Yes\\
  % 0.5 &K,$\kappa$ & - & -  & 160,150,130,110 & 3,3,3,3 & & 15,000 & 30,000 & 10 & 240x120x128& no\\
0.5 &K, nonthermal & - & -  & 160 & 3 &$4.1 \times 10^{-9}$ & 15,000 & 30,000 & 10 & $240\times120\times128$& No\\
0.5 &K, nonthermal & - & -  & 150 & 3 &$4.1 \times 10^{-9}$ & 15,000 & 30,000 & 10 & $240\times120\times128$& No\\
0.5 &K, nonthermal & - & -  & 130 & 3 &$4.1 \times 10^{-9}$ & 15,000 & 30,000 & 10 & $240\times120\times128$& No\\
0.5 &K, nonthermal & - & -  & 110 & 3 &$4.1 \times 10^{-9}$ & 15,000 & 30,000 & 10 & $240\times120\times128$& No\\
   %% \hline lower fluxes, just checking
   %% 0 & K & - & - & 160,150,130,110 & 3.5,3.5,3.5,3 & 15,000 & 30,000 & 10 \\
   %% 0.5 &K & - & -  & 160,150,130,110 & 2,2,2,2 & 15,000 & 30,000 & 10 \\
   %% 0.9375&K & - & - & 160,150,130,110 & 1.5,1.5,1.5,1.5  & 15,000 & 30,000 & 10 \\
   \hline
  \end{tabular}
  \caption{List of GRMHD/GRRMHD and radiative transfer models resulting in different
    time series of images. Note. Accretion rates $\dot{M}$ are time averages between $t_{\rm s}$ and $t_{\rm f}$.}\label{tab:all}
\end{center}
\end{table*}

\clearpage

%% fig 1 

\clearpage

%% fig2

\clearpage

%% fig3

\clearpage

%% fig4

%% fig5

%\clearpage

%fig6

\clearpage

% fig7 done

\clearpage

% fig8 done

\clearpage

%% fig10

\clearpage

%% fig11

%\clearpage

%% fig12

\clearpage

\appendix

\section{Light curves}\label{app:lc}

\begin{figure*}
  \centering
  \includegraphics[width=1.0\linewidth]{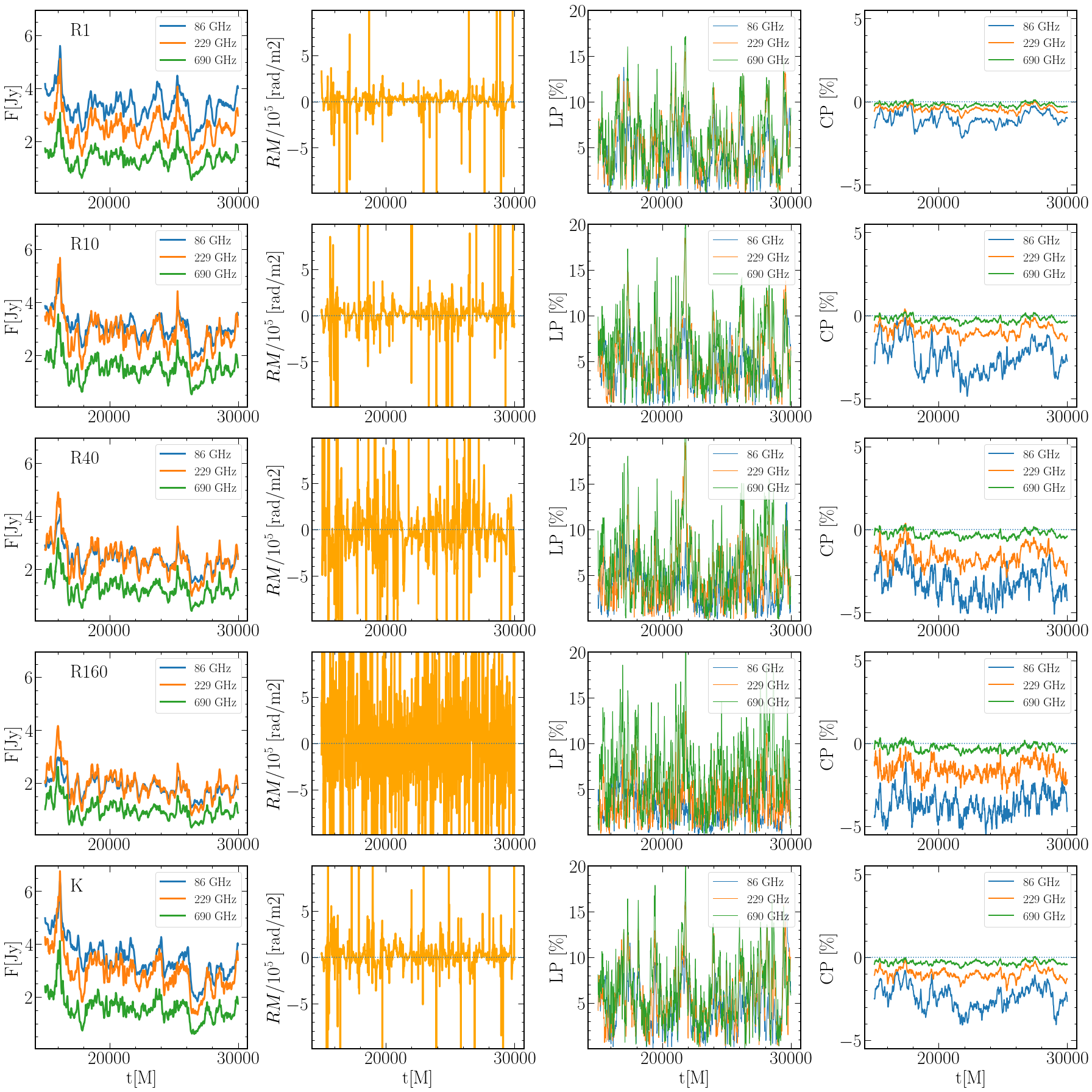}
\caption{Radiative signatures for the MAD $a_* = 0$ model at the default viewing
  angle of $i = 160^\circ$. Panels from left to right display Stokes ${\mathcal
    I}$ (86, 229, 690 GHz), RM (213 - 229 GHz), LP (86, 229, 690 GHz) and CP
  (86, 229, 690 GHz). Panels from top to bottom show models R1-160, and K.}\label{fig:lc_MADa0} 
\end{figure*}

\begin{figure*}
  \centering
  \includegraphics[width=1\linewidth]{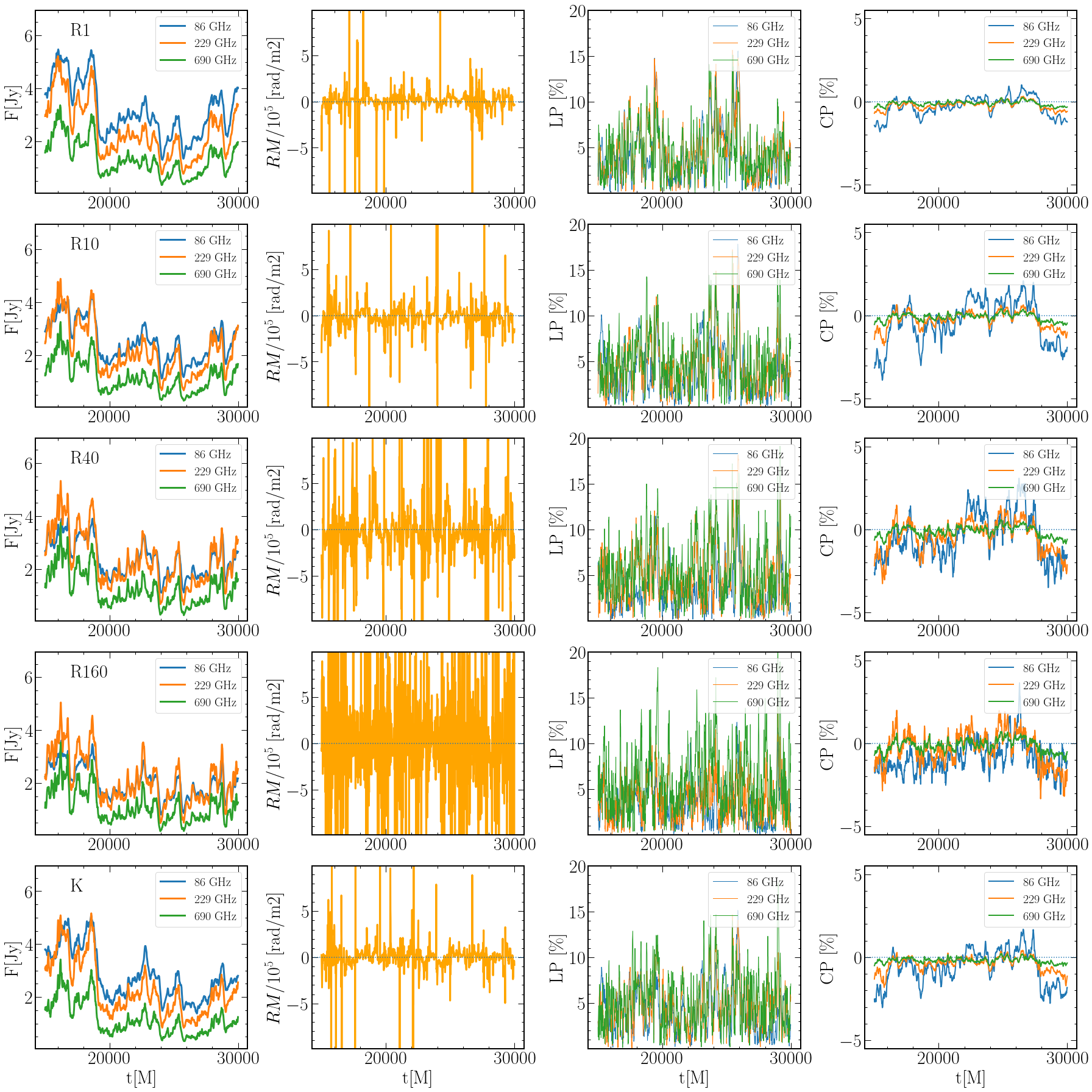}
\caption{Same as in Figure~\ref{fig:lc_MADa0} but for MAD $a_* = 0.5$
  models R1 - 160 and K.}\label{fig:lc_MADa5} 
\end{figure*}

\begin{figure*}
  \centering
  \includegraphics[width=1\linewidth]{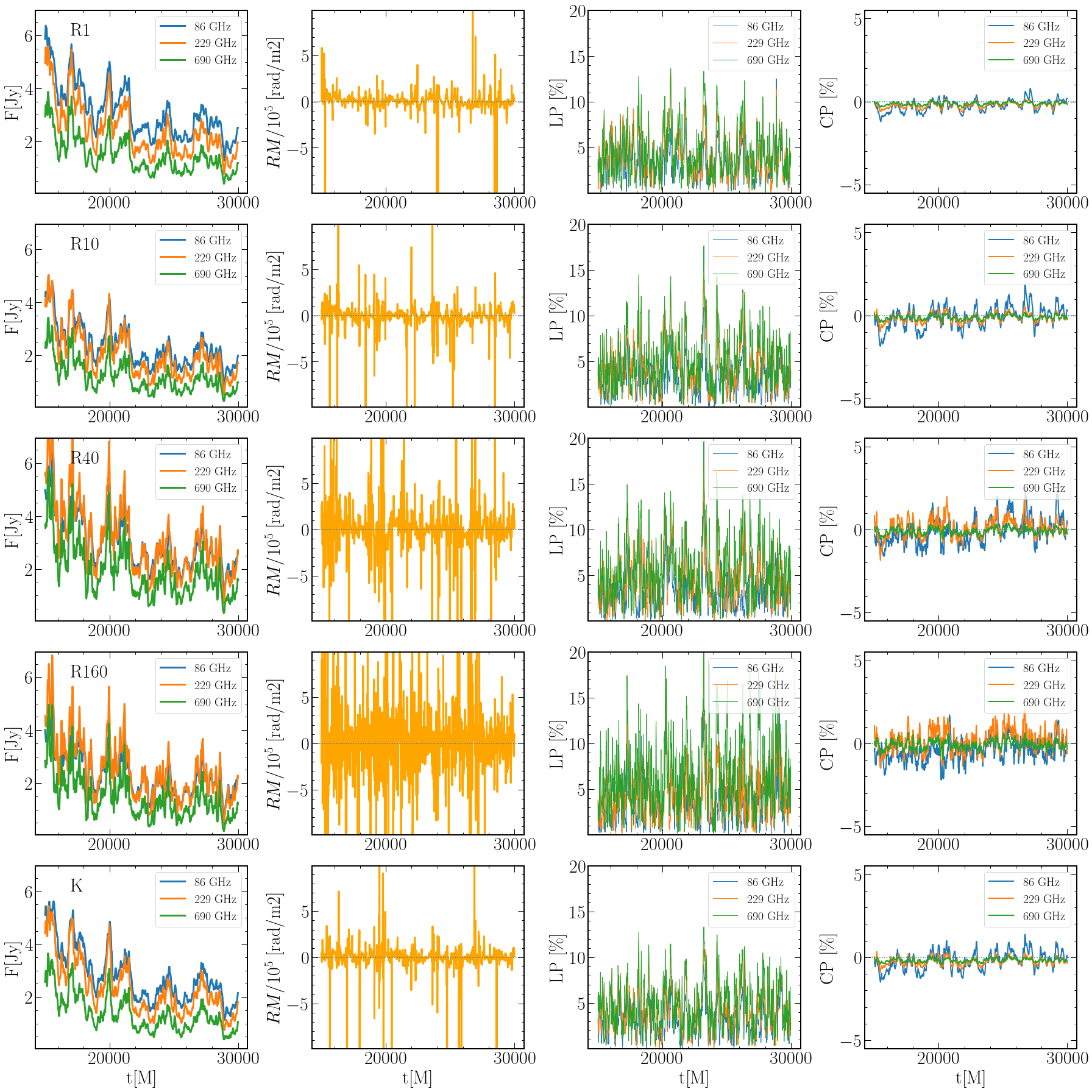}
\caption{Same as in Figure~\ref{fig:lc_MADa0} but for MAD $a_* = 0.94$
  models R1 - 160 and K.}\label{fig:lc_MADa9} 
\end{figure*}

The main text of this manuscript presents statistical properties of
light curves from MAD simulations with different spin values. 
Figures~\ref{fig:lc_MADa0} - \ref{fig:lc_MADa9} present light curves of
individual zero and prograde models for the default viewing angle $i = 160^\circ$.
Here in addition to 86 and 229 GHz we also show light curves at 690 GHz for visual comparison.

\clearpage

%% fig13

%% fig14

%% fig15

\clearpage

\section{GRRMHD}\label{app:grrmhd}

In \texttt{ebhlight}, the multifrequency radiative transport is carried out using the
Monte Carlo scheme in which the radiation field is simulated with a large
number of superphoton particles representing an even larger number of physical photons (see code description in \citealt{ryan:2015} for exact definitions). To make sure that the radiative cooling of plasma
is well captured, for each fluid element, the Monte Carlo scheme has to produce a
sufficient number of photon particles within the electron cooling timescale.

\begin{figure}
  \centering
  \includegraphics[width=1\linewidth]{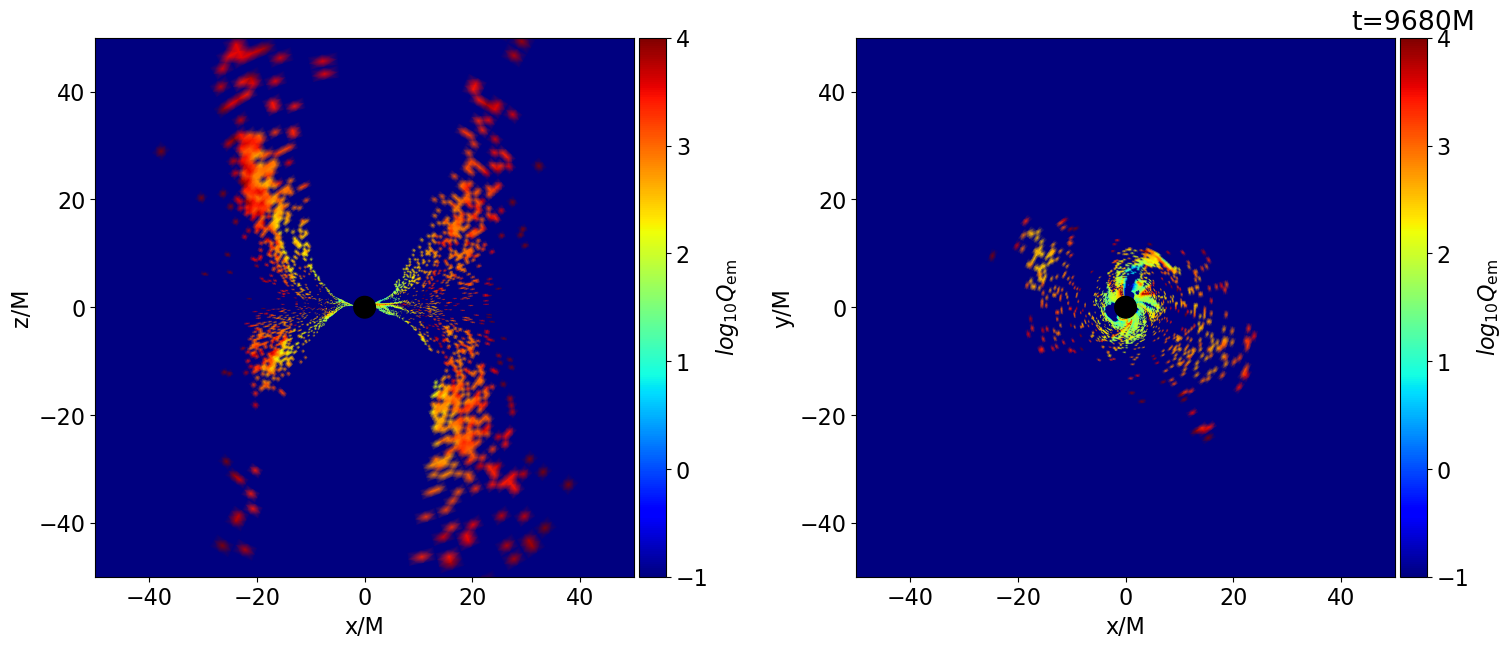}\\
\caption{Meridional and equatorial slices showing number of superphotons
  emitted within one synchrotron cooling timescale in GRRMHD MAD $a_* = 0$
  model. The figure illustrates not only the quality of the GRRMHD model but
  also the origin of the synchrotron emission in the two-temperature MAD models.}\label{fig:Qem}
\end{figure}

The electron cooling timescale (in M units) can be written as 
\begin{equation}
\tau_{\rm cool} = \frac{ u_{\rm e, code}}{\Lambda_{\rm code}} 
\end{equation}
where $u_{\rm e,code}$ is the electron internal energy provided by the two-temperature
model (in code units) and where $\Lambda_{\rm code}$ is the synchrotron
cooling rate (also in code units). In cgs units the cooling rate reads (Eq.~(A4) from \citealt{moscibrodzka:2011})
\begin{equation}
\Lambda_{\rm c.g.s.}=\frac{16 B^2e^4 n_{\rm e} \Theta_e^2}{3c^3m_{\rm e}^2}.
\end{equation}  
where the conversion to code units is $\Lambda_{\rm code}=\Lambda_{\rm cgs} {\mathcal L} {\mathcal T^3}/{\mathcal M}$.
Figure~\ref{fig:Qem} shows the quality factor of GRRMHD $Q_{\rm em}$, a number of superphotons
emitted within $\tau_{\rm cool}$, for a snapshot of GRRMHD around
$t$ = 10,000M. $Q_{\rm em} \gg 10$ in the inner disk ($r<10$M) and along the jet wall, indicating
reasonable sampling in regions where radiative cooling is the most efficient.

% fig 16

\clearpage

\section{The issue of $R_{\rm low}$ parameter in $R~(\beta)$ model and parameter inference}\label{app:rlow}

\begin{figure}
  \centering
  \includegraphics[width=1.0\linewidth]{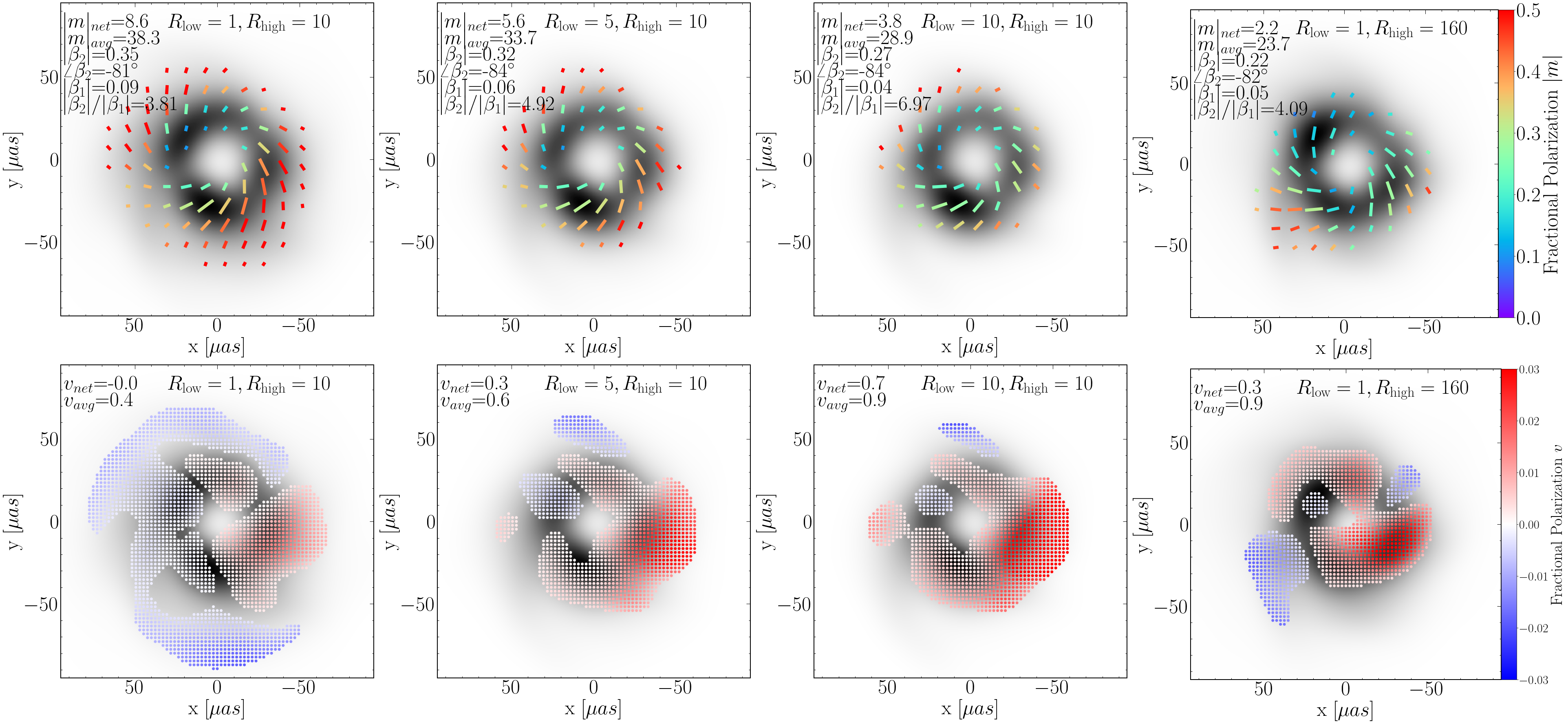}
  \caption{Polarimetric images of a single GRMHD MAD $a_* = 0.5$ snapshot. Panels show the same quantities as in Figure~\ref{fig:resolved_pol}
  but only for $R~(\beta)$ models with different assumptions of $R_{\rm low}$
  parameter in Equation~\ref{eq:rhigh}. All models are normalized with
  different ${\mathcal M}$ to reproduce the same total flux. From left to
  right: ${\mathcal M}/10^{17}$ = 3, 4, 5, 8.}\label{fig:rlow}
\end{figure}

When estimating the parameters of the electron temperature using polarimetric images of Sgr~A*, \citet{EHTC:2024VIII} assumed
that in Equation~\ref{eq:rhigh}, $R_{\rm low} = 1$. 
What happens if we relax the assumption about $R_{\rm low}$? Figure~\ref{fig:rlow} shows the impact of
the varying parameter $R_{\rm low}$ on the linear and circular polarimetric images of Sgr~A*.
The images of the MAD with ($R_{\rm low}, R_{\rm high})=(10, 10)$ start to
resemble those of ($R_{\rm low}, R_{\rm high})=(1, 160)$. Therefore, it is possible that the polarimetric scoring of the models with $R_{\rm low} \sim 10$ could lead
to different best-bet values of $R_{\rm high} \ll 160$. If this is indeed the
case, which will have to be carefully verified in the future, estimations of
the electron temperature parameters using $R~(\beta)$ models with $R_{\rm low}
\sim 10$ have a better chance of convergence with some of the discussed \qratio models, proving that particle-in-cell models applied to GRMHD simulations correctly predict the order of magnitude of electron heating in collisionless plasma. 

%fig 17

\end{document}